\documentclass[twocolumn]{aastex62}

\newcommand{\itbold}[1]{\textbf{\textit{#1}}}
\newcommand{\pdif}[2]{\frac{\partial #1}{\partial #2}}
\newcommand{\odif}[2]{\frac{d #1}{d #2}}

\graphicspath{{./}{figures/}{./data/pk/}{./data/vpk/}{./data/snapshot/}{./data/mass_func/}}

\received{}
\revised{}
\accepted{\today}

\submitjournal{ApJ}

\shorttitle{Vlasov simulation of Cosmic Relic Neutrinos}
\shortauthors{Yoshikawa et al.}

\usepackage{amsmath}	
\usepackage{multirow}
\begin{document}

\title{Cosmological Vlasov--Poisson Simulations of Structure Formation with Relic Neutrinos: Nonlinear Clustering and the Neutrino Mass}

\correspondingauthor{Kohji Yoshikawa}
\email{kohji@ccs.tsukuba.ac.jp}

\author[0000-0003-0389-5551]{Kohji Yoshikawa}
\affiliation{Center for Computational Sciences, University of Tsukuba, 1-1-1 Tennodai, Tsukuba, Ibaraki 305-8577, Japan}

\author{Satoshi Tanaka}
\affiliation{Yukawa Institute for Theoretical Physics, Kyoto University, Kitashirakawa Oiwake-Cho, Sakyo-Ku, Kyoto 606-8502, Japan}

\author{Naoki Yoshida}
\affiliation{Department of Physics, The University of Tokyo, Bunkyo, Tokyo 113-0033, Japan}
\affiliation{Kavli Institute for the Physics and Mathematics of the
Universe, The University of Tokyo, Kashiwa, Chiba 277-8583, Japan}
\affiliation{Research Center for the Early Universe, The University of Tokyo, Bunkyo, Tokyo 113-0033, Japan}

\author{Shun Saito}
\affiliation{Institute for Multi-messenger Astrophysics and Cosmology, Department of Physics, Missouri University of Science and Technology, 1315 N Pine St, Rolla, MO 65409}
\affiliation{Kavli Institute for the Physics and Mathematics of the
Universe, The University of Tokyo, Kashiwa, Chiba 277-8583, Japan}

\begin{abstract}
 We present the results of cosmological simulations of large-scale
 structure formation with massive neutrinos.  The phase-space
 distribution of the cosmic relic neutrinos is followed, for the first
 time, by directly integrating the six-dimensional Vlasov--Poisson
 equations.  Our novel approach allows us to represent free streaming
 and clustering of neutrinos, and their gravitational interaction with
 cold dark matter accurately. We thus obtain solutions for the
 collisionless dynamics independent of conventional $N$-body methods.  We
 perform a suite of hybrid $N$-body/Vlasov simulations with varying
 the neutrino mass, and systematically examine the dynamical effects of
 massive neutrinos on large-scale structure formation.  Our simulations
 show characteristic large-scale clustering of the neutrinos and their
 coherent streaming motions relative to dark matter.  The effective
 local neutrino ``temperature'' around massive galaxy clusters varies by
 several percent with respect to the cosmic mean; the neutrinos in
 clusters can be hotter or colder depending on the neutrino mass.  We
 study a number of statistics of the large-scale structure and of dark
 matter halos in comparison with those obtained by $N$-body
 simulations and/or by perturbation theory. Our simulations mark an
 important milestone in numerical cosmology, and pave a new way to
 study cosmic structure formation with massive neutrinos.
\end{abstract}

\keywords{large-scale structure of universe --- dark matter --- neutrinos --- methods: numerical}

\section{Introduction}
A generic prediction of the standard Big Bang cosmology is the existence
of a relic neutrino background that pervades the Universe.  Neutrinos
had been long considered to be massless in the standard model of
particle physics, but the experimental confirmation of neutrino flavor
oscillation \citep{Fukuda1998} suggests that at least two of the three
kinds of neutrinos have non-zero masses. However, their absolute masses
and the hierarchy among the three mass eigenstates remain unknown, and
precise measurement of the neutrino mass by terrestrial experiments is
still challenging.

The dynamical influence of the permeating relic neutrinos is imprinted
in the large-scale mass distribution in the Universe.  The major effect
of massive neutrinos is suppression of clustering of dark matter and
hence of galaxies.  Collisionless damping caused by fast-moving
neutrinos effectively ``drags'' the growth of structure, which can be
discerned, for instance, as suppression of the galaxy power spectrum at
smaller scale than the neutrinos free streaming scale.  The free
streaming scale and the degree of damping depend on the absolute mass of
neutrinos (see \citet{Lesgourgues2006} for a detailed review).  Thus, it
is possible to constrain or even {\it measure} the absolute mass of
neutrinos and their mass hierarchy from observations of the LSS in the
universe
\citep[e.g.,][]{Hu1998,Takada2006,Saito2009a,Namikawa2010,Font-Ribera2014,Boyle2018,Chudaykin2019}.

The most stringent constraints on the absolute mass of neutrinos are
given by observations of the CMB anisotropies \citep{Planck2018}.  The
total mass of three mass eigenstates of neutrinos is estimated to be
smaller than $0.26\,{\rm eV}\,(95\%\,{\rm C.L.})$ with the Planck data
alone, and a tighter constraint of $0.12\,{\rm eV}$ is derived from the
combination with the CMB lensing as well as with the baryon acoustic
oscillation (BAO) data.

It is of crucial importance to provide an accurate theoretical
prediction of the dynamical effect of the relic neutrinos particularly
in nonlinear regime, in order to place tight constraint on the absolute
neutrino masses.  To this end, a number of authors use perturbation
theory and/or attempt analytical modelling of the evolution of density
fluctuations under the presence of massive neutrinos
\citep[e.g.,][]{Saito2008, Wong2008, Shoji2009, Lesgourgues2009,
Blas:2014, Dupuy2015, Fuehrer2015, Peloso2015, Levi2016, Senatore:2017}.
\citet{Ichiki2012} investigate the dynamical effect of massive neutrinos
on non-linear spherical collapse of cold dark matter (CDM) halos
starting from a top-hat CDM overdensity. They show that massive
neutrinos effectively suppress the mass function of dark matter (DM)
halos even if the total neutrino mass is lightest possible inferred from
the neutrino oscillation experiments, 0.05 eV or 0.1 eV for the normal
and inverted mass hierarchy, respectively.  Despite of the theoretical
challenges, the LSS observables are shown to derive competitive
constraints on the total neutrino masses
\citep[e.g.,][]{Seljak2006,Ichiki2009,Thomas2010,Saito2011,Zhao2013,Beutler2014,Alam2017,Palanque-Delabrouille2020,Ivanov2020,Aviles2020}.

Direct numerical simulations have been used to study the dynamical
impact of massive neutrinos on the LSS formation
\citep[e.g.,][]{Brandbyge2009, Brandbyge2010a, Brandbyge2010b, Viel2010,
Bird2012, Ali-Haimoud2013, Villaescusa-Navarro2014, Upadhye2014,
Castoria2014, Inman2015, Inman2017, Banerjee2016, Wright2017,
McCarthy2018, Banerjee2018, Villaescusa-Navarro2019}.  Many of these
simulations employ particle-based techniques and attempt to reproduce
free-streaming of massive neutrinos by adding thermal velocities
randomly sampled according to the velocity distribution function of
massive neutrinos.  The validity of such treatment is unclear, however.
Also the particle-based simulation method generically suffers from
numerical shot noise owing to relatively coarse sampling of neutrino
distribution in the six-dimensional phase space \citep{YYU2013}.  Poor
sampling in the velocity space leads to inaccurate treatment of the
free-streaming, because the high velocity tails of the distribution
function, a dynamically crucial component, is not smoothly represented.

As a possible remedy to reduce the particle shot noise, grid-based
approaches \citep{Brandbyge2009, Viel2010} and a hybrid method with a
particle-based one \citep{Brandbyge2010a} are proposed. In the latter,
the density field of massive neutrinos is {\it assumed} to evolve
according to linear theory.  Unfortunately, these approximate methods do
not follow non-linear evolution of neutrino dynamics and the
gravitational interaction with the CDM component consistently. Recently,
\citet{Inman2015, Inman2017} performed $N$-body simulations employing an
extremely large number of particles to reduce the shot noise
contamination. While employing a large number of particles is a
straightforward way to mitigate discreteness effects, it would be
desirable to devise and use a method that can accurately represents the
velocity distribution function and follow its time evolution.

In this paper, we present a novel approach that directly follows the
time evolution of the distribution function of neutrinos in
six-dimensional phase space. We directly solve the collisionless
Boltzmann equation, or the Vlasov equation, using a finite-volume
method. Since our approach treats massive neutrinos as a continuous
collisionless {\it fluid}, it is not compromised by numerical shot
noise. This approach, which is referred to as Vlasov simulation
hereafter, was first applied to numerical simulations of
self-gravitating systems by \citet{Fujiwara1981, Fujiwara1983a}, and
also applied to DM halo formation with massive neutrinos
\citep{Fujiwara1983b}, although limited in one or two spatial
dimensions.  A number of Vlasov solvers have been developed with
employing various methods and have been applied to plasma and
self-gravitating systems \citep{Filbet2003, Colombi2014, Colombi2017}.
The simulations of \citet{YYU2013} are the first of the kind that are
performed with three spatial dimensions (in six-dimensional phase
space).  The Vlasov simulations are able to accurately reproduce
free-streaming of a collisionless fluid, an effect expected to be of
critical importance in following the dynamics of cosmological relic
neutrinos. In the present paper, we run a series of Vlasov simulations
of the LSS formation with massive neutrinos, and study the clustering
and the dynamical effect of neutrinos systematically with varying the
total neutrino mass.

The rest of the paper is organized as follows. We describe our new
numerical method presented in this work in Section~\ref{sec:method}.
Section~\ref{sec:run} and \ref{sec:results} are devoted to describe
numerical simulations and their results regarding the dynamical effect
of massive neutrinos on the LSS formation. In Section~\ref{sec:cost}, we
address the computational cost of our Vlasov simulation and the
conventional $N$-body simulations in terms of the amount of memory
requirement, computational wall clock time, and the spatial resolution
of neutrinos' distribution. Finally, we summarize our work in
Section~\ref{sec:summary}.

\bigskip

\section{Formulation and Numerical Method}
\label{sec:method}

We adopt a novel numerical scheme that follows the distribution of the
neutrinos by directly integrating the Vlasov equation in the full
six-dimensional phase space. For the CDM component, we perform
high-resolution $N$-body simulations because CDM is assumed to have a
very small thermal velocity dispersion with a very compact distribution
in velocity space.  The efficient hybrid $N$-body/Vlasov simulations
allow us to reproduce the nonlinear clustering of CDM and neutrinos in a
self-consistent manner. In this section, we lay out basic theoretical
formulation for the dynamics of CDM and neutrinos in a cosmological
context.

\subsection{Cosmological relic neutrinos}

In the hot primeval plasma in the early universe, neutrinos are in
thermal equilibrium with radiation and with other matter, and have a
Fermi-Dirac distribution. After the neutrinos decouple from other
matter, the distribution function freezes out because the neutrinos have
little interaction with other component afterwards. Neutrinos with
non-zero masses become non-relativistic at some early epoch, and then
the velocity distribution function is given by
\begin{equation}
 \label{eq:FermiDirac}
 F_{\rm FD}(\itbold{v}, t) = \left[\exp\left(\frac{cm_\nu|\itbold{v}|}{k_{\rm B}T_\nu}\right)+1\right]^{-1},
\end{equation}
where $m_\nu$ is a mass eigenvalue of neutrinos, $c$ the speed of light,
$k_{\rm B}$ the Boltzmann constant, $T_\nu$ the redshifted mean
"temperature", and $\itbold{v}$ is the peculiar velocity of
neutrinos. Here, we do not consider the chemical potential of neutrinos
because it is thought to be negligible.  The temperature of the cosmic
neutrino background is proportional to, and slightly lower than, that of
the cosmic microwave background (CMB) photons $T_{\rm CMB}$ as
\begin{equation}
 T_\nu = \left(\frac{4}{11}\right)^{1/3} T_{\rm CMB}.
\end{equation}
The numerical factor originates from photon heating during a brief
period of electron-positron annihilation.  The comoving number density
of neutrinos, $n_\nu$, can be obtained by
integrating~(\ref{eq:FermiDirac}) over the entire velocity space, and
the density parameter of non-relativistic neutrinos at the present
universe is given by
\begin{equation}
 \Omega_\nu = \sum_i n_\nu m_\nu^i c^2/\rho_{\rm c}=\frac{M_\nu c^2}{93.14\,h^2{\rm eV}},
\end{equation}
where $\rho_{\rm c}$ is the critical energy density, $h$ is the present
value of the Hubble parameter in units of 100 km s$^{-1}$ Mpc$^{-1}$,
$m_\nu^i$ is the neutrino mass of the $i$-th mass eigenstate, and
$M_\nu$ is the sum of three mass eigenvalues of neutrinos.

\subsection{Vlasov simulation of neutrinos}
The Vlasov equation describing the dynamics of neutrinos in the
cosmological comoving frame is given by
\begin{equation}
 \label{eq:comoving_vlasov}
  \pdif{f}{t} + \frac{\itbold{u}}{a^2}\cdot\pdif{f}{\itbold{x}}-\pdif{\phi}{\itbold{x}}\cdot\pdif{f}{\itbold{u}} = 0
\end{equation}
where $f(\itbold{x},\itbold{u},t)$ is the distribution function of
neutrinos, $\phi(\itbold{x})$ is the gravitational potential, $a(t)$ is
the scale factor of cosmological expansion, $\itbold{x}$ is the comoving
coordinate and $\itbold{u}=a^2\dot{\itbold{x}}$ is the canonical
velocity. The canonical velocity is convenient as a ``velocity''
variable in our Vlasov simulations, because the neutrino peculiar
velocity $a\dot{\itbold{x}}$ decay as $a(t)^{-1}$ in the limit of the
uniform and homogeneous universe. Then the canonical velocity remains
roughly constant in the linear evolution phase. Note also that the bulk
velocity of the neutrinos is typically much smaller than, or only
comparable to, the velocity dispersion. These features are numerically
convenient because the extent of the velocity distribution in the
initial condition does not vary significantly in terms of canonical
velocity. We refer the readers to Appendix~\ref{sec:velocitycoordinate}
for more detailed discussion on the choice of velocity coordinate such
as peculiar velocity.

The distribution function is normalized so that the integration over the
velocity space yields the mass density {\it contrast} as
\begin{equation}
 \label{eq:distribution_function_normalization}
  \int f(\itbold{x},\itbold{u},t)\, d^3\itbold{u} = 1+\delta_{\nu}(\itbold{x},t),
\end{equation}
where $\delta_\nu(\itbold{x},t)$ is the mass density fluctuation of the
neutrinos.

As in \citet{YYU2013}, the six-dimensional phase space is discretized
onto uniform cartesian mesh grids (hereafter Vlasov mesh grids) both in
the physical and velocity spaces in a finite volume manner. The number
of mesh grids is referred to as $N_{\rm x}$ for the physical space, and
$N_{\rm u}$ for the velocity space, respectively. We configure cubic
physical and velocity domains with $0 \le x, y, z \le L_{\rm box}$ and
$-u_{\rm max} \le u_x, u_y, u_z \le u_{\rm max}$.

We adopt the directional splitting method to solve
equation~(\ref{eq:comoving_vlasov}).  We effectively solve six
one-dimensional advection equations as
\begin{equation}
 \label{eq:advection_in_physical_space}
 \pdif{f}{t} + \frac{u_i}{a^2}\pdif{f}{x_i} = 0\,\,\,\,\,(i=1,2,3)
\end{equation}
and
\begin{equation}
 \label{eq:advection_in_velocity_space}
 \pdif{f}{t} - \pdif{\phi}{x_i}\pdif{f}{u_i} = 0\,\,\,\,\,(i=1,2,3),
\end{equation}
where $(x_1, x_2, x_3)=(x,y, z)$ and $(u_1, u_2, u_3)=(u_x, u_y, u_z)$.
These advection equations are numerically solved with SL-MPP7 scheme
\citep{Tanaka2017}, which has spatially seventh-order accuracy and
ensures the monotonicity and positivity of the numerical solution.

We adopt outflow boundary conditions in the velocity space. If a
fraction of neutrinos are accelerated beyond the predefined velocity
boundaries, they are simply treated as disappeared.  Our simulation code
automatically detects such an unphysical situation and monitors the
conservation of total mass of neutrinos.  In practice, we configure a
large enough velocity space and the total loss from the velocity
boundaries is sufficiently small and the fractional error of mass
conservation is less than $0.001\%$.  We impose periodic boundary
conditions on the physical space as is often adopted in cosmological
simulations.

\subsection{$N$-body Simulation of CDM}

We employ the conventional $N$-body method for CDM.  We assume that the
simulation particles represent baryons and CDM, and do not treat them
separately nor consider the hydrodynamic effect of baryons.  Hereafter,
the combined component of CDM and baryons is referred to as ``CDM'' for
simplicity.

The motion of each particle is determined by the equation of motion
\begin{equation}
 \label{eq:eom}
 \odif{^2 \itbold{x}}{t^2} + 2 H \odif{\itbold{x}}{t} = -\frac{\nabla\phi}{a^2},
\end{equation}
where $H(t)\equiv\dot{a}/a$ is the Hubble parameter. We adopt a leapfrog
integrator as described in Section 2.5. We compute the gravitational
forces on $N$-body particles with Particle-Mesh (PM)
scheme~\citep{Hockney1981}. The Poisson solver is modified suitably to
incorporate the gravitational interaction of neutrinos with CDM.

\subsection{Gravitational Potential}
 
The gravitational potential in Equation~(\ref{eq:eom}) is the same as in
Equation~(\ref{eq:comoving_vlasov}). Thus both CDM and neutrinos share
the common gravitational field, which satisfies the Poisson equation
\begin{equation}
 \label{eq:poisson}
 \nabla^2\phi = 4\pi
  G\bar{\rho}_{\rm m}(t)a^2\delta_{\rm m},
\end{equation}
where $\bar{\rho}_{\rm m}(t)$ and $\delta_{\rm m}$ are the mean mass
density and its fluctuation of the total matter composed of CDM and
neutrinos.

For a given distribution of CDM particles, we compute the density field
using the Triangular Shaped Cloud (TSC) scheme on a uniform mesh grid
(hereafter the PM mesh) with $N_{\rm PM}$ grid points. The density field
of neutrinos is computed at $N_{\rm x}$ spatial grid points employed in
the Vlasov simulation.  We integrate (sum) the discretized distribution
function over the velocity space.  Since $N_{\rm PM}$ is set to be
larger than $N_{\rm x}$, the total matter (CDM + neutrinos) density is
obtained by up-sampling the neutrino density field by $N_{\rm PM}/N_{\rm
x}$-fold and then by adding the contribution to the CDM density. The
Poisson equation~(\ref{eq:poisson}) is numerically solved with the
convolution method \citep{Hockney1981} that applies fast Fourier
transform to the total matter density field.  The gradient of the
gravitational potential is evaluated by the four-point finite difference
approximation (FDA), and is interpolated at positions of $N$-body
particles with the TSC scheme. When integrating the Vlasov equation, the
gravitational potential on the PM mesh is down-sampled onto the Vlasov
mesh, and the gravitational force in Equation~(\ref{eq:comoving_vlasov})
is computed with the six-point FDA scheme.

\subsection{Time Integration}

The time step width for integrating Equations~(\ref{eq:comoving_vlasov})
and (\ref{eq:eom}) is constrained by the conditions in the $N$-body
simulation for CDM and in the Vlasov simulation for neutrinos.  In the
$N$-body simulation, the time step $\Delta t_{\rm N}$ is determined as
\begin{equation}
 \Delta t_{\rm N} = \min\left(\min_i\left(\frac{\Delta_{\rm PM}}{|\dot{\itbold{x}}_i|}\right) ,\min_i\left(\sqrt{\frac{\Delta_{\rm PM}}{|\nabla\phi_i|/a^2}}\right)\right),
\end{equation}
where $\nabla\phi_i$ is the gradient of gravitational potential at the
position of $i$-th $N$-body particle, $\Delta_{\rm PM}=L_{\rm
box}/N_{\rm PM}^{1/3}$ the grid spacing of the PM mesh.  We search the
global minimum by considering the above two criteria for all the
$N$-body particles.

In the Vlasov simulation, the time step $\Delta t_{\rm V}$ is restricted
by the CFL condition
\begin{equation}
 \Delta t_{\rm V} = \nu_{\rm CFL}\min\left(\frac{\Delta_{\rm x}}{u_{\rm max}/a^2}, \min_j\left(\frac{\Delta_{\rm u}}{|\phi_{x,j}|}, \frac{\Delta_{\rm u}}{|\phi_{y,j}|}, \frac{\Delta_{\rm u}}{|\phi_{z,j}|}\right)\right),
\end{equation}
where $\phi_{x,j}$, $\phi_{y,j}$ and $\phi_{z,j}$ represent numerical
partial derivatives of gravitational potential $\phi$ with respect to
$x$, $y$ and $z$ at the $j$-th mesh grid for the physical space, and
$\Delta_{\rm x}=L_{\rm box}/N_{\rm x}^{1/3}$ and $\Delta_{\rm u}=2u_{\rm
max}/N_{\rm u}^{1/3}$ are the physical and velocity spacings of the
Vlasov mesh grids, respectively.  The CFL parameter $\nu_{\rm CFL}$ is
set to $\nu_{\rm CFL}=0.2$ as suggested in \citet{Tanaka2017}.

Equations~(\ref{eq:comoving_vlasov}) and (\ref{eq:eom}) are numerically
integrated simultaneously with a time step given by
\begin{equation}
 \Delta t = \min(\Delta t_{\rm N}, \Delta t_{\rm V}), 
\end{equation}
in a Kick--Drift--Kick (KDK) leapfrog manner \citep{Quinn1997,
Springel2005}.
To integrate the system over a single time step $t^n$ to
$t^{n+1}=t^n+\Delta t$, we perform a number of steps sequentially in the
following order:
\begin{enumerate}
 \item We first compute the gravitational potential field $\phi^n$ using
       the positions of the CDM particles and the distribution function
       of neutrinos at $t=t^n$.

 \item Comoving velocities of CDM particles $\dot{\itbold{x}}^n$ are
       updated to $\dot{\itbold{x}}^{n+1/2}$ by a half of the time step
       as
       \begin{equation}
	\label{eq:leapfrog1}
	\begin{split}	 
	\dot{\itbold{x}}^{n+1/2} &= \dot{\itbold{x}}^n \frac{1-H^n\cdot\Delta t/2}{1+H^n\cdot\Delta t/2} \\ 
	 &-\frac{\nabla\phi^n}{(a^n)^2}\frac{1}{1+H^n\cdot\Delta t/2},
	\end{split}
       \end{equation}
       where $a^n$ and $H^n$ are the scale factor and the Hubble
       parameter at $t=t^n$, respectively.
       
 \item The distribution function of the neutrinos $f^n(\itbold{x},
       \itbold{u})$ at $t=t^n$ is advected in the velocity space using
       the gravitational potential $\phi^n$ at $t=t^n$ by sequentially
       integrating Equations~(\ref{eq:advection_in_velocity_space}) over
       $\Delta t/2$, to yield an updated distribution function
       $f^*(\itbold{x},\itbold{u})$ formally given by
       \begin{equation}
	f^*(\itbold{x},\itbold{u})=f^n(\itbold{x}, \itbold{u}+\nabla\phi^n\cdot\Delta t/2).
       \end{equation}

 \item The positions of the CDM particles $\itbold{x}^n$ are evolved to
       $\itbold{x}^{n+1}$ by a full time step of $\Delta t$ using the
       comoving velocity $\dot{\itbold{x}}^{n+1/2}$ as
       \begin{equation}
	\itbold{x}^{n+1} = \itbold{x}^n + \dot{\itbold{x}}^{n+1/2}\cdot\Delta t.
       \end{equation}
       
 \item The neutrino distribution function is advected in the physical
       space by numerically integrating the advection
       equations~(\ref{eq:advection_in_physical_space}) by a time step
       of $\Delta t$. In this step, we set $a^{n+1/2}=a(t^n+\Delta t/2)$
       for the scale factor in
       Equations~(\ref{eq:advection_in_physical_space}). The updated
       distribution function $f^{**}(\itbold{x}, \itbold{u})$ is
       formally written as
       \begin{equation}
	f^{**}(\itbold{x}, \itbold{u}) = f^*(\itbold{x}-\itbold{u}/(a^{n+1/2})^2\cdot\Delta t, \itbold{u}).
       \end{equation}

 \item Gravitational potential $\phi^{n+1}$ at $t=t^{n+1}$ is computed
       using the updated positions of the CDM particles and the neutrino
       distribution function $f^{**}(\itbold{x}, \itbold{u})$ obtained
       in the previous steps. It should be noted that the density field
       of neutrinos obtained from the distribution function
       $f^{**}(\itbold{x},\itbold{u})$ is identical to that fromp
       $f^{n+1}(\itbold{x},\itbold{u})$ computed in the next procedure.

 \item Comoving velocities of CDM particles are evolved from
       $\dot{\itbold{x}}^{n+1/2}$ to $\dot{\itbold{x}}^{n+1}$ with
       $\phi^{n+1}$ by a time step of $\Delta t/2$ as
       \begin{equation}
	\label{eq:leapfrog2}
	\begin{split}
	 \dot{\itbold{x}}^{n+1} &= \dot{\itbold{x}}^{n+1/2} \frac{1-H^{n+1}\cdot\Delta t/2}{1+H^{n+1}\cdot\Delta t/2} \\ 
	 & -\frac{\nabla\phi^{n+1}}{(a^{n+1})^2}\frac{1}{1+H^{n+1}\cdot\Delta t/2}, 
	\end{split}
       \end{equation}
       and the distribution function of neutrinos $f^{**}(\itbold{x},
       \itbold{u})$ is advected in the physical space by sequentially
       integrating Equations~(\ref{eq:advection_in_physical_space})
       using the gravitational potential $\phi^{n+1}$ at $t=t^{n+1}$ to
       obtain the distribution function $f^{n+1}(\itbold{x},\itbold{u})$
       at $t=t^{n+1}$ written as
       \begin{equation}
	f^{n+1}(\itbold{x}, \itbold{u}) = f^{**}(\itbold{x},\itbold{u}+\nabla\phi^{n+1}\cdot\Delta t/2)
       \end{equation}
\end{enumerate}

\begin{figure}[htbp]
 \centering
 \includegraphics[width=8.5cm]{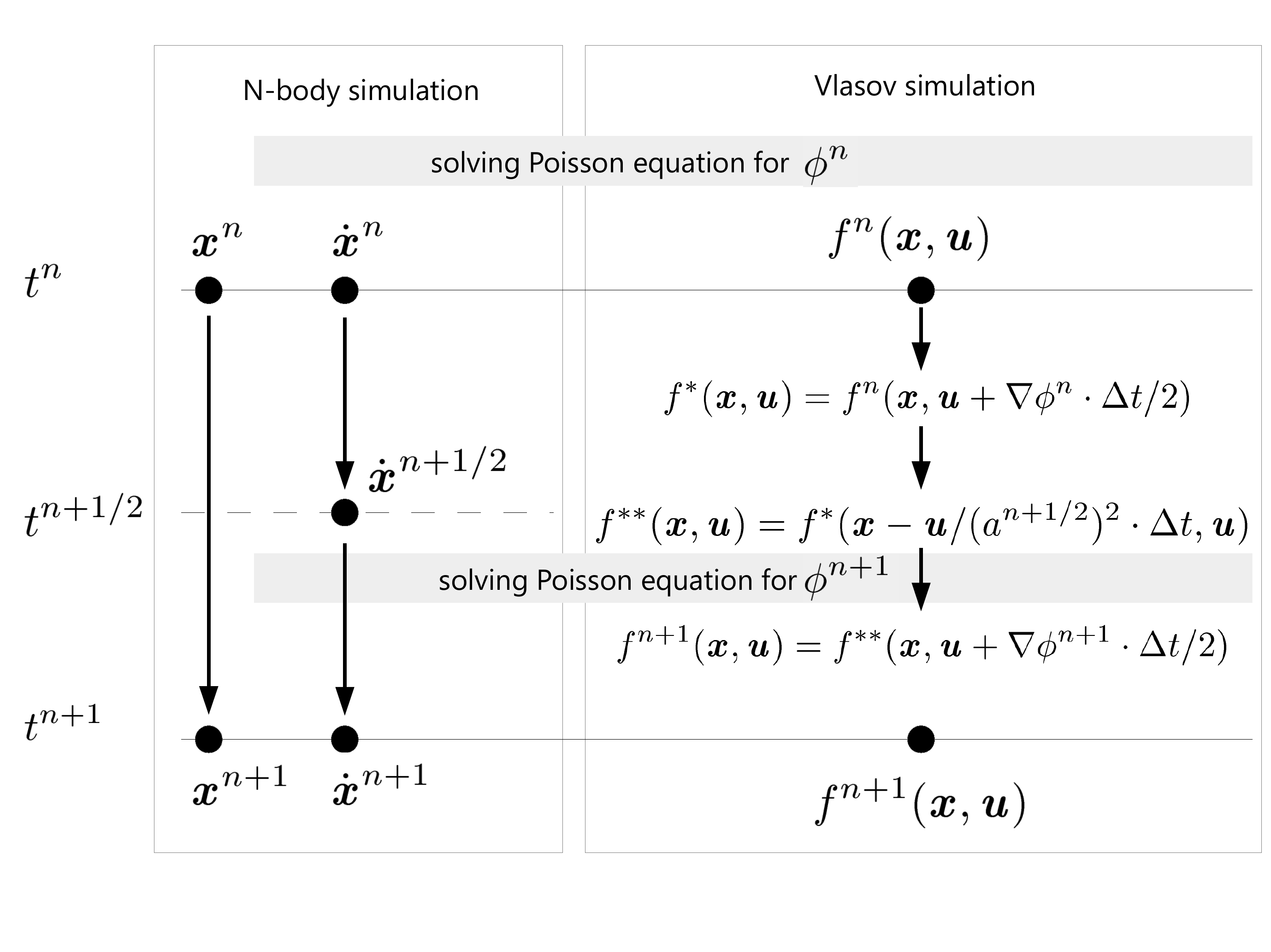}

 \figcaption{Schematic description of the time integration scheme of our
 hybrid $N$-body/Vlasov simulation.\label{fig:integration}}
\end{figure}

Figure~\ref{fig:integration} shows a schematic description of numerical
procedures in integrating $N$-body simulation and Vlasov simulation
simultaneously by a single time step.

\bigskip

\section{Cosmological Simulations}
\label{sec:run}

\subsection{Models}
We assume a spatially flat cosmology with the present matter density
parameter $\Omega_{\rm m}=0.308$, and that of baryons $\Omega_{\rm
b}=0.0484$, the cosmological constant $\Omega_{\rm v} = 0.692$, and the
hubble constant $h=0.678$. The model is consistent with the CMB
observation by Planck satellite \citep{Planck2015XIII}.

We consider five cases with the total rest mass energy of three neutrino
mass eigenstates of 0 eV (massless neutrinos), 0.1 eV, 0.2 eV, 0.3 eV
and 0.4 eV. The present-epoch density parameters of neutrinos and CDM
are given by
\begin{equation}
 \Omega_\nu = 9.34\times 10^{-3} \left(\frac{M_\nu c^2}{0.4\, {\rm eV}}\right).
\end{equation}
and $\Omega_{\rm c} = \Omega_{\rm m}-\Omega_\nu$, respectively.
Table~\ref{tbl:models} summarize the set of numerical simulations.  We
run simulations with different volumes with the comoving sidelength
$L_{\rm box}$ ranging from 200 $h^{-1}$Mpc to 10 $h^{-1}$Gpc. We set the
number of Vlasov mesh grids to $(N_{\rm x}, N_{\rm u}) = (128^3,64^3)$,
and the number of CDM particles to $N_{\rm N}=1024^3$. We perform
convergence tests by comparing the simulations with different box sizes
(Section 4.2). The mass of a single CDM particle is
\begin{equation}
 m_{\rm CDM} = 9.39\times 10^8 f_{\rm CDM} \left(\frac{L_{\rm box}}{200h^{-1}{\rm Mpc}}\right)^3M_{\odot},
\end{equation}
where $f_{\rm CDM}$ is the CDM mass fraction given by $f_{\rm CDM} =
\Omega_{\rm c}/\Omega_{\rm m}$.

We assume that the masses of the three mass eigenstates are equal to
each other.  In practice, we follow the time evolution of only one
distribution function because neutrinos in the three mass eigenstates
can be treated identically under this assumption. We note that this is
indeed a good approximation in the cases with $M_\nu c^2 \gtrsim 0.2\,\,
{\rm eV}$ \citep{Lesgourgues2006}.  However, if the total mass is
smaller, i.e. $M_\nu c^2 \lesssim 0.1\,\, {\rm eV}$, the kinematic
properties of the neutrinos cannot be represented accurately by a single
distribution function.  In the present study, we nevertheless assume an
equal mass of neutrinos among the three eigenstates to avoid additional
memory consumption necessary to handle multiple distribution functions
in six-dimensional phase space.  The results of our simulations with
$M_\nu c^2 =0.1\,\,{\rm eV}$ should thus be regarded as being less
accurate compared to the other cases in terms of the dynamical effect of
massive neutrinos on the LSS formation.

\bigskip

\begin{table}[h]
 \centering
 \begin{tabular}{|c|c|c|c|c|}
  \tableline
  $L_{\rm box}$ & $M_\nu c^2$ [eV] & $\Omega_{\rm c}$ & $N_{\rm run}$ & $\sigma_8$\\
  \tableline
  200$h^{-1}$Mpc & \multirow{4}{*}{0.0}& \multirow{4}{*}{0.308}& 4 & \multirow{4}{*}{0.819}\\
  600$h^{-1}$Mpc &  & & 2 &\\
  1$h^{-1}$Gpc &  & & 1 &\\
  10$h^{-1}$Gpc &  & & 1 & \\
  \tableline
  200$h^{-1}$Mpc & \multirow{4}{*}{0.1} & \multirow{4}{*}{0.306} & 4 & \multirow{4}{*}{0.804}\\
  600$h^{-1}$Mpc &  & & 2 &\\
  1$h^{-1}$Gpc &  & & 1 &\\
  10$h^{-1}$Gpc &  & & 1 & \\
  \tableline
  200$h^{-1}$Mpc & \multirow{4}{*}{0.2} & \multirow{4}{*}{0.303} & 4 &\multirow{4}{*}{0.785}\\
  600$h^{-1}$Mpc &  & & 2 &\\
  1$h^{-1}$Gpc &  & & 1 &\\
  10$h^{-1}$Gpc &  & & 1 & \\
  \tableline
  200$h^{-1}$Mpc & \multirow{4}{*}{0.3}& \multirow{4}{*}{0.301} & 4 &\multirow{4}{*}{0.765}\\
  600$h^{-1}$Mpc & & & 2 &\\
  1$h^{-1}$Gpc & & & 1 &\\
  10$h^{-1}$Gpc & & & 1 & \\
  \tableline
  200$h^{-1}$Mpc & \multirow{4}{*}{0.4} & \multirow{4}{*}{0.299} & 4 &\multirow{4}{*}{0.745}\\
  600$h^{-1}$Mpc &  & & 2 &\\
  1$h^{-1}$Gpc & & & 1 &\\
  10 $h^{-1}$Gpc & & & 1 & \\
  \tableline
 \end{tabular}
 \caption{Numerical models simulated in this work. $N_{\rm run}$ is the
 number of realizations.  \label{tbl:models}}
\end{table}

\subsection{Initial Conditions}
We generate the initial conditions at the redshift of $z_{\rm ini} = 10$
for all our simulations listed in Table~\ref{tbl:models}.  We first
compute the initial power spectra of density fluctuations of CDM and
neutrinos separately using the {\tt CAMB} software package
\citep{Lewis2000} for the cosmological parameters given in Section 3.1.
The spectral index of the primordial density fluctuations is set to be
$n_{\rm s}=0.96$, and we normalize the fluctuation amplitudes such that
the amplitude of the curvature perturbation power spectrum is
$\Delta_{\cal R} = 2.37\times 10^{-9}$ at the pivot scale of $k_{\rm p}
= 0.002 \,\,{\rm Mpc}^{-1}$.

The initial positions of the CDM particles are determined by displacing
the particles from a regular ``lattice'' distribution. The velocities
are computed by using the Zel'dovich approximation. More accurate
initial conditions can be generated by using a method based on
second-order Lagrangian perturbation theory (2LPT) \citep{Crocce2006}.
Although our method without the second-order correction might yield
slightly inaccurate power spectra of the initial density fluctuations at
small length scales \citep{Nishimichi2009}, we expect that the overall
impact to our simulations is limited, because strong nonlinear
clustering of CDM is driven gravitationally at late epochs through to
$z=0$. We do not incorporate the effect of scale-dependent growth factor
and growth rate in computing velocity fields of CDM and massive
neutrinos \citep{Zennaro2017}, which is expected to be a minor impact on
our initial conditions since the initial redshift is relatively low.  In
our future work, we will perform detailed comparison and convergence
tests using multiple methods for the CDM initial condition.  In the
present paper, we focus on relative differences in various statistical
quantities between the cases with relic neutrinos with different total
masses (Section 4).

For neutrinos, we assume the initial distribution function at a time
$t_i$ is given by the Fermi--Dirac distribution as
\begin{equation}
 f(\itbold{x}, \itbold{u}, t_i) = \frac{1+\delta_\nu(\itbold{x})}{N}F_{\rm FD}((\itbold{u}-\itbold{u}_{\rm b})/a(t_i), t_i)
\end{equation}
where $\itbold{u}_{\rm b}(\itbold{x})$ is the canonical bulk velocity of
neutrinos at the location $\itbold{x}$, and the normalization factor
\begin{equation}
 N = \int F_{\rm FD}((\itbold{u}-\itbold{u}_{\rm b}(\itbold{x}))/a(t_i),t_i)d^3\itbold{u}
\end{equation}
is introduced to satisfy the normalization of the neutrinos'
distribution function given by
Equation~(\ref{eq:distribution_function_normalization}).  We do not
consider the small distortion of the phase space distribution induced
during the relativistic and trans-relativistic phases
\citep{Ma1994}. The extent of velocity space $u_{\rm max}$ is set as
$u_{\rm max} = 4\,\sigma_{\rm u}$ where $\sigma_{\rm u}$ is the
dispersion of canonical velocity given by
\begin{equation}
 \sigma_{\rm u}^2 = \frac{\displaystyle\int u^2 f(\itbold{x},\itbold{u},t_i)d^3\itbold{u}}{\displaystyle\int f(\itbold{x},\itbold{u},t_i)d^3\itbold{u}},
\end{equation}
and is approximated as
\begin{equation}
 \label{eq:velc_disp}
 \sigma_{\rm u} \simeq 180 \left(\frac{m_\nu c^2}{1\, {\rm eV}}\right)^{-1}\,\,{\rm km/s}.
\end{equation}
Note that the peculiar velocity dispersion is given by
\begin{equation}
 \label{eq:velc_disp2}
 \sigma_{\rm v} = 180 (1+z)\left(\frac{m_\nu c^2}{1\, {\rm eV}}\right)^{-1}\,\,{\rm km/s}.
\end{equation}
In practice, our simulations show that the estimated $u_{\rm max}$ is
sufficiently large to enclose the velocity extent of the distribution
function of neutrinos even after evolved to $z=0$. As suggested by
\citet{YYU2013}, the velocity resolution $\Delta_{\rm u}$ should satisfy
the condition
\begin{equation}
 \label{eq:vel_resolution}
 \sigma_u/\Delta_{\rm u} \gtrsim 5 
\end{equation}
to reproduce the neutrino free streaming accurately. The choice of
$u_{\rm max} = 4 \sigma_{\rm u}$ yields the velocity resolution of
$\Delta_{\rm u} = 2u_{\rm max}/N_{\rm u}^{1/3}=\sigma_{\rm u}/8$
satisfying the above condition.

\subsection{Computational Facility}

Our $N$-body/Vlasov hybrid simulations presented in this work are
performed on the Oakforest-PACS (OfP) supercomputer installed in Joint
Center for Advanced High Performance Computing
(JCAHPC)\footnote{http://jcahpc.jp/eng/index.html}. Each computing node
consists of a Intel Xeon Phi (Knights Landing) processor, 96 GiB DDR4
RAM and 16 GiB of high-bandwidth MCDRAM. The simulations listed in
Table~\ref{tbl:models} are run on 512 nodes, and typically consume 15
wallclock hours for the runs with $L=200h^{-1}\,{\rm Mpc}$.

\bigskip

\section{Results}
\label{sec:results}

\subsection{The Distribution of Neutrinos}

\begin{figure*}[htbp]
 \centering

 \includegraphics[width=17cm]{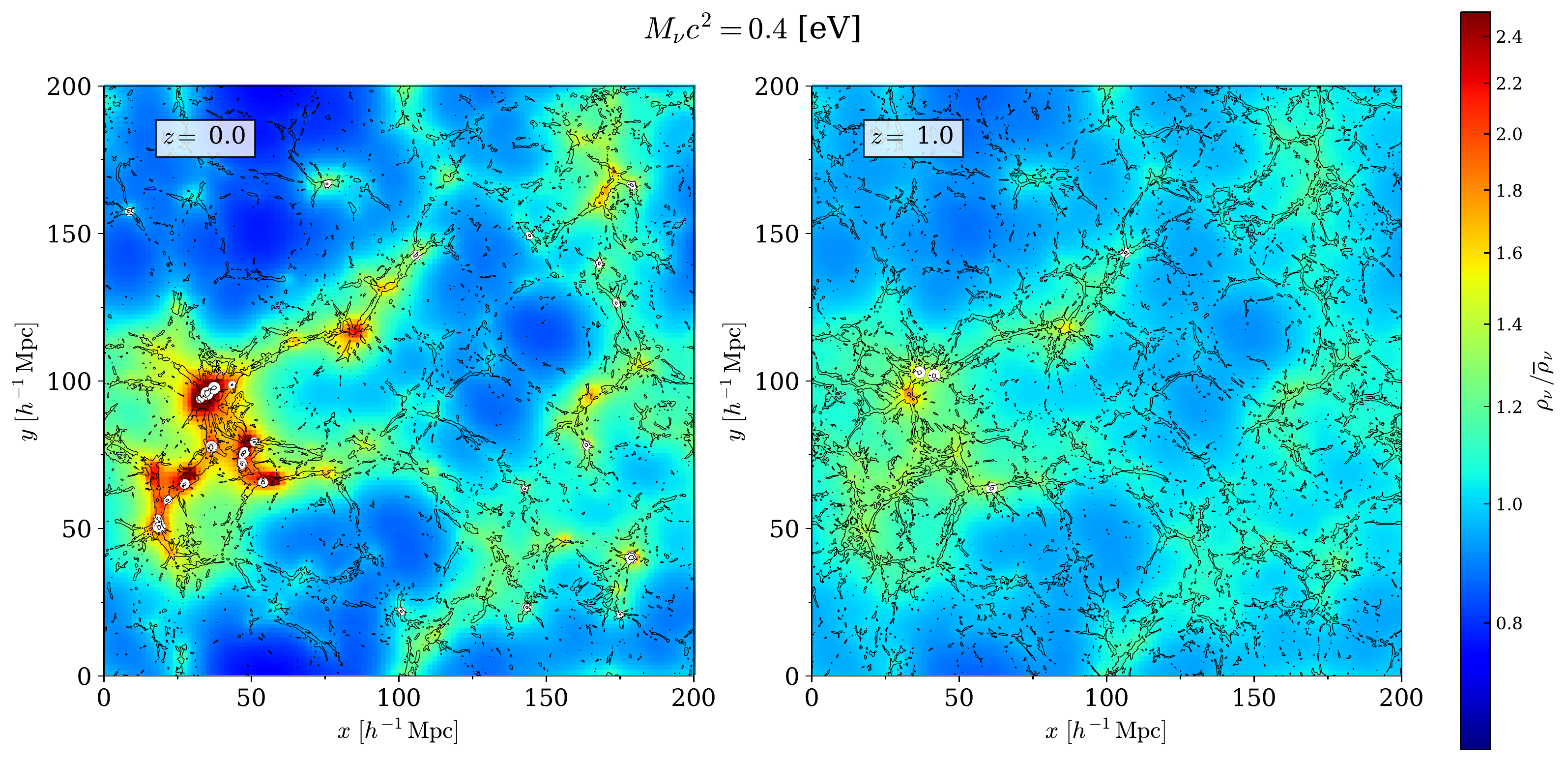}
 \includegraphics[width=17cm]{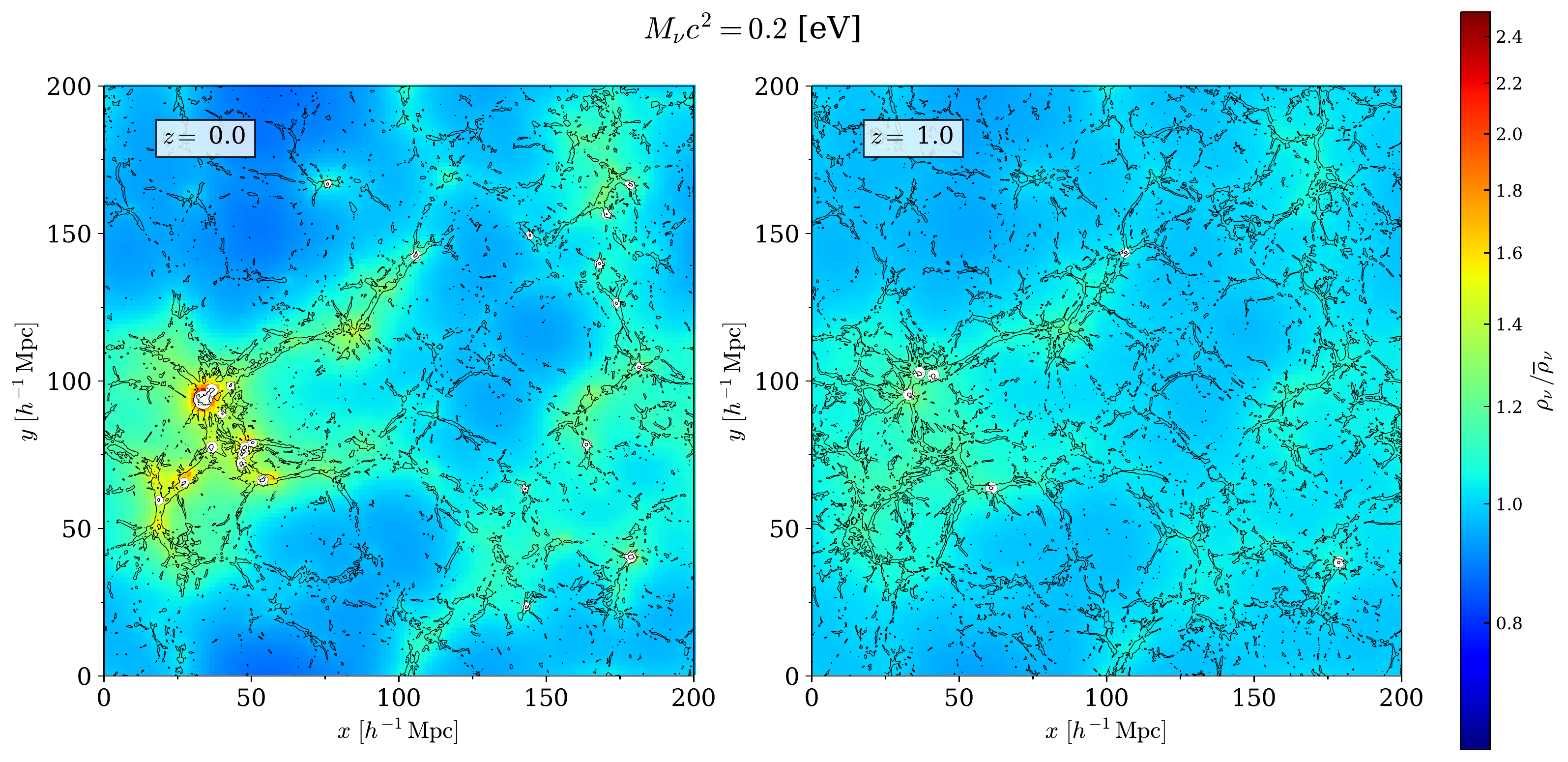}

 \figcaption{Density fields of CDM (contour) and neutrinos (color) in
 the runs with $L_{\rm box}=200h^{-1}\,\,{\rm Mpc}$, $M_\nu
 c^2=0.4\,\,{\rm eV}$ (upper panels) and 0.2 eV (lower panels) at
 $z=0.0$ (left) and 1.0 (right). Colorbars indicate the mass overdensity
 of neutrinos. Circles indicate positions of DM halos with a virial mass
 greater than $10^{13} M_{\odot}$. The radii of the circles are
 proportional to the virial radii of DM halos.
 \label{fig:snapshot_dens}}
\end{figure*}

Figure~\ref{fig:snapshot_dens} shows the density distributions of CDM
and neutrinos, as well as the distribution of massive DM halos with
virial mass $M >10^{13}{{M_\odot}}$, in the runs with $L_{\rm
box}=200h^{-1}$ Mpc.  The plotted region is a slice with a width of
$\Delta_{\rm x}$, and the contours and color maps represent the
overdensity of CDM and neutrinos, respectively.  The region presented in
Figure~\ref{fig:snapshot_dens} contains the most massive DM halo with
mass of $1.26\times 10^{15} M_{\odot}$ and $9.84 \times 10^{14}
M_{\odot}$ in the runs with $M_\nu c^2 =0.2$ eV and 0.4 eV, respectively
at $z=0$. The distribution of neutrinos is much smoother than that of
CDM owing to their large velocity dispersion. The neutrino density
contrast is smaller for the smaller neutrino mass, but the maximum
over-density reaches $\delta_{\nu} \sim 2$ at $z=0$; the large-scale
nonlinear clustering is clearly seen even for the light neutrino model
with $M_\nu c^2 = 0.2$ eV.

\begin{figure*}[htbp]
 \centering 
 \includegraphics[width=17cm]{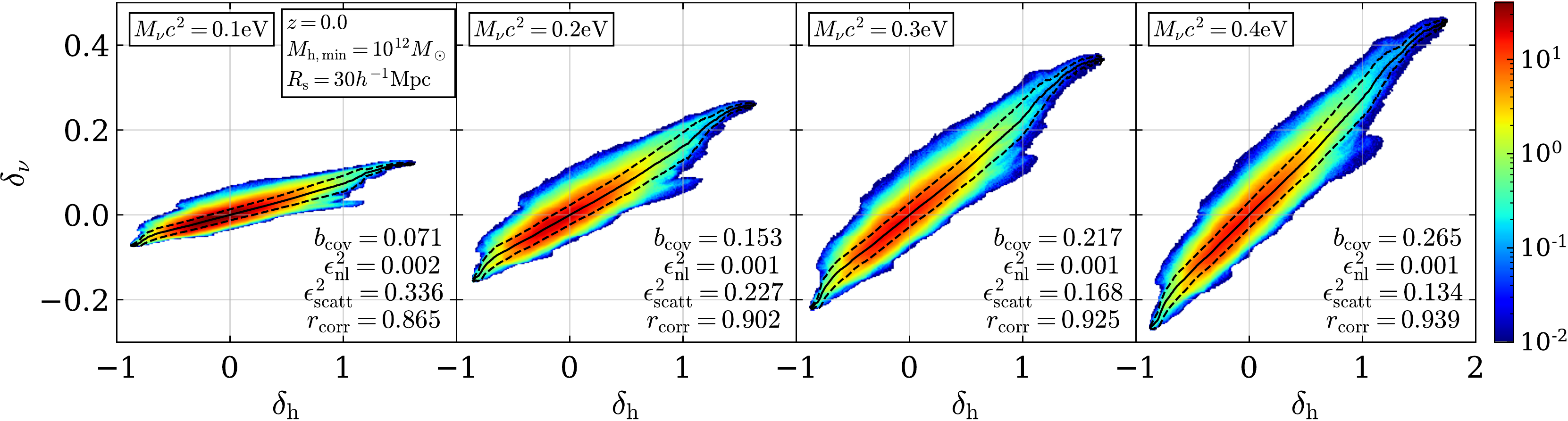}
 \includegraphics[width=17cm]{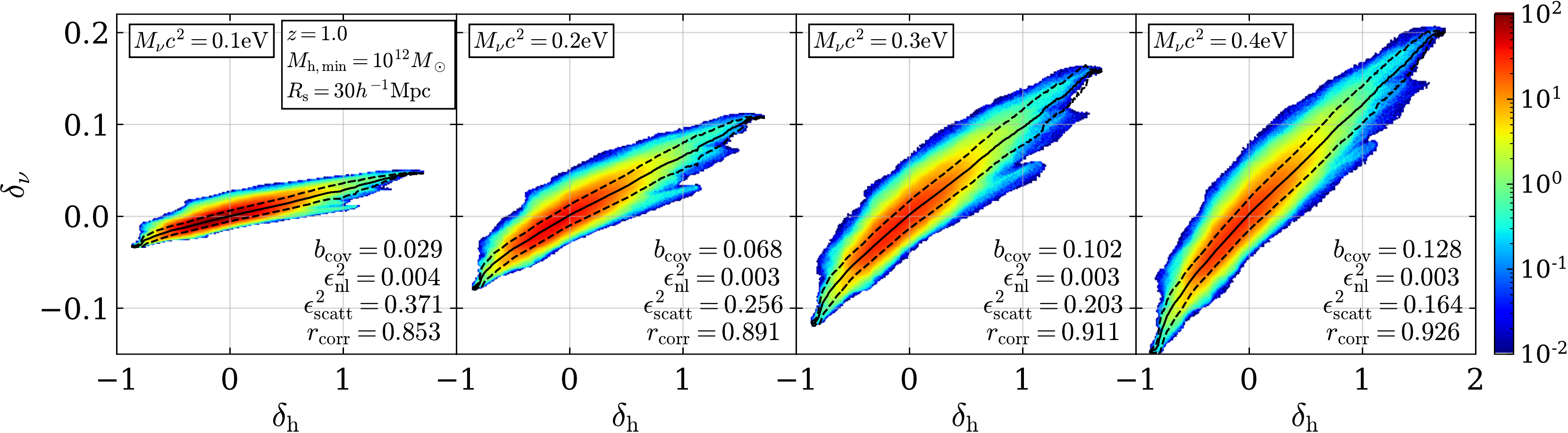}

 \figcaption{Probability distribution function of the number density of
 DM halos $\delta_{\rm h}$ and the mass density of neutrinos
 $\delta_\nu$ at $z=0$ and $z=1$. Both the quantities are smoothed with
 a smoothing scale of $R_s=30h^{-1}\,{\rm Mpc}$.  We consider DM halos
 with virial mass greater than $10^{12}M_\odot$. In each panel, the
 solid line indicates the mean value of neutrino density contrast for a
 given DM halo overdensity, and two dotted lines indicate the 1-$\sigma$
 dispersion of $\delta_\nu$ around the mean. \label{fig:bias}}
\end{figure*}

Interestingly, the local neutrino density differs significantly in
regions with similar CDM densities. We also find massive DM halos tend
to be located in neutrino-rich regions.  In order to study this tendency
further, we compute a joint probability distribution of the mass density
fluctuation of neutrinos $\delta_\nu$ and the number density
fluctuations of DM halos defined as
\begin{equation}
 \delta_{\rm h}(\itbold{x}) = \frac{n_{\rm h}(\itbold{x})-\bar{n}_{\rm h}}{\bar{n}_{\rm h}},
\end{equation}
where $n_{\rm h}(\itbold{x})$ is the number density of DM halos with
virial mass greater than a certain threshold $M_{\rm h,min}$, and
$\bar{n}_{\rm h}$ is the mean halo number density. Figure~\ref{fig:bias}
shows the joint probability distribution of $\delta_{\rm h}$ and
$\delta_\nu$, where $M_{\rm h,min}$ is set to $10^{12}\,M_\odot$ and
both of $\delta_{\rm h}$ and $\delta_\nu$ are smoothed over a comoving
scale of $R_{\rm s}=30h^{-1}\, {\rm Mpc}$ with a top-hat filter.  We
also compute the mean relation of the neutrino density fluctuation
$\overline{\delta}_\nu(\delta_{\rm h})$ and its variance
$\overline{\delta^2_\nu}(\delta_{\rm h})$ as a function of $\delta_{\rm
h}$ given by
\begin{equation}
 \overline{\delta}_\nu(\delta_{\rm h}) = \int P(\delta_{\rm h}, \delta_\nu)\delta_\nu d\delta_\nu,
\end{equation}
and
\begin{equation}
 \overline{\delta^2_\nu}(\delta_{\rm h}) = \int P(\delta_{\rm h}, \delta_\nu) \delta_\nu^2 \,d\delta_\nu
\end{equation}
where $P(\delta_{\rm h},\delta_\nu)$ is the joint probability
distribution function. $\overline{\delta}_\nu(\delta_{\rm h})$ and
$\overline{\delta}_\nu(\delta_{\rm
h})\pm\left[\overline{\delta^2_\nu}(\delta_{\rm h})\right]^{1/2}$ are
plotted by solid and dotted lines, respectively, in each panel. We
numerically derive several parameters that characterize the bias of
neutrino mass density fluctuation relative to the number density
fluctuation of DM halos following the nonlinear and stochastic model of
\citet{Taruya2000}.  The neutrino bias is defined as
\begin{equation}
 b_{\rm cov} = \frac{\langle\delta_\nu\delta_{\rm h}\rangle}{\langle\delta_{\rm h}^2\rangle},
\end{equation}
and the nonlinearity and the stochasticity are given by, respectively,
\begin{equation}
 \epsilon_{\rm nl}^2=\frac{\langle\delta_\nu^2\rangle\langle\overline{\delta}_\nu^2\rangle}{\langle\delta_\nu\delta_{\rm h}\rangle^2}-1,
\end{equation}
and
\begin{equation}
 \epsilon_{\rm scatt}^2 = \frac{\langle\delta_{\rm h}^2\rangle\langle(\delta_\nu-\overline{\delta}_\nu)^2\rangle}{\langle\delta_\nu\delta_{\rm h}\rangle^2}.
\end{equation}
We show the measured values of these quantities in each panel of
Figure~\ref{fig:bias}. We also compute the correlation coefficient
between $\delta_\nu$ and $\delta_{\rm h}$ defined as
\begin{equation}
 r_{\rm corr} = \frac{\langle\delta_\nu\delta_{\rm h}\rangle}{\sqrt{\langle\delta_\nu^2\rangle\langle\delta_{\rm h}^2\rangle}}.
\end{equation}
Both $\epsilon_{\rm nl}^2$ and $\epsilon_{\rm scatt}^2$ are
significantly smaller than unity, and hence the correlation coefficient,
which scales as $r_{\rm corr} = (1+\epsilon_{\rm nl}^2+\epsilon_{\rm
scatt}^2)^{-1/2}$, is close to unity regardless of the neutrino mass,
but the bias $b_{\rm cov}$ is smaller for less massive neutrinos.  We
note that these features are similar to the biasing relation between DM
density field and DM halo number density studied by \citet{Taruya2000}
and \citet{Yoshikawa2001}.

The mean $\overline{\delta}_\nu - \delta_{\rm h}$ relations are almost
linear with small dispersions, suggesting that the local neutrino
density can be inferred from the number density field of DM halos
smoothed over a certain length scale for the cases with $M_\nu c^2\geq
0.2\,{\rm eV}$ , where we find $\overline{\delta}_\nu(\delta_{\rm h})
\gg \langle \overline{\delta^2}_\nu(\delta_{\rm h})\rangle ^{1/2}$
around $\delta_{\rm h}\simeq 1$.

\bigskip

\subsection{Velocity Dispersion of Neutrinos}

\begin{figure*}[htbp]
 \centering
 \includegraphics[width=17cm]{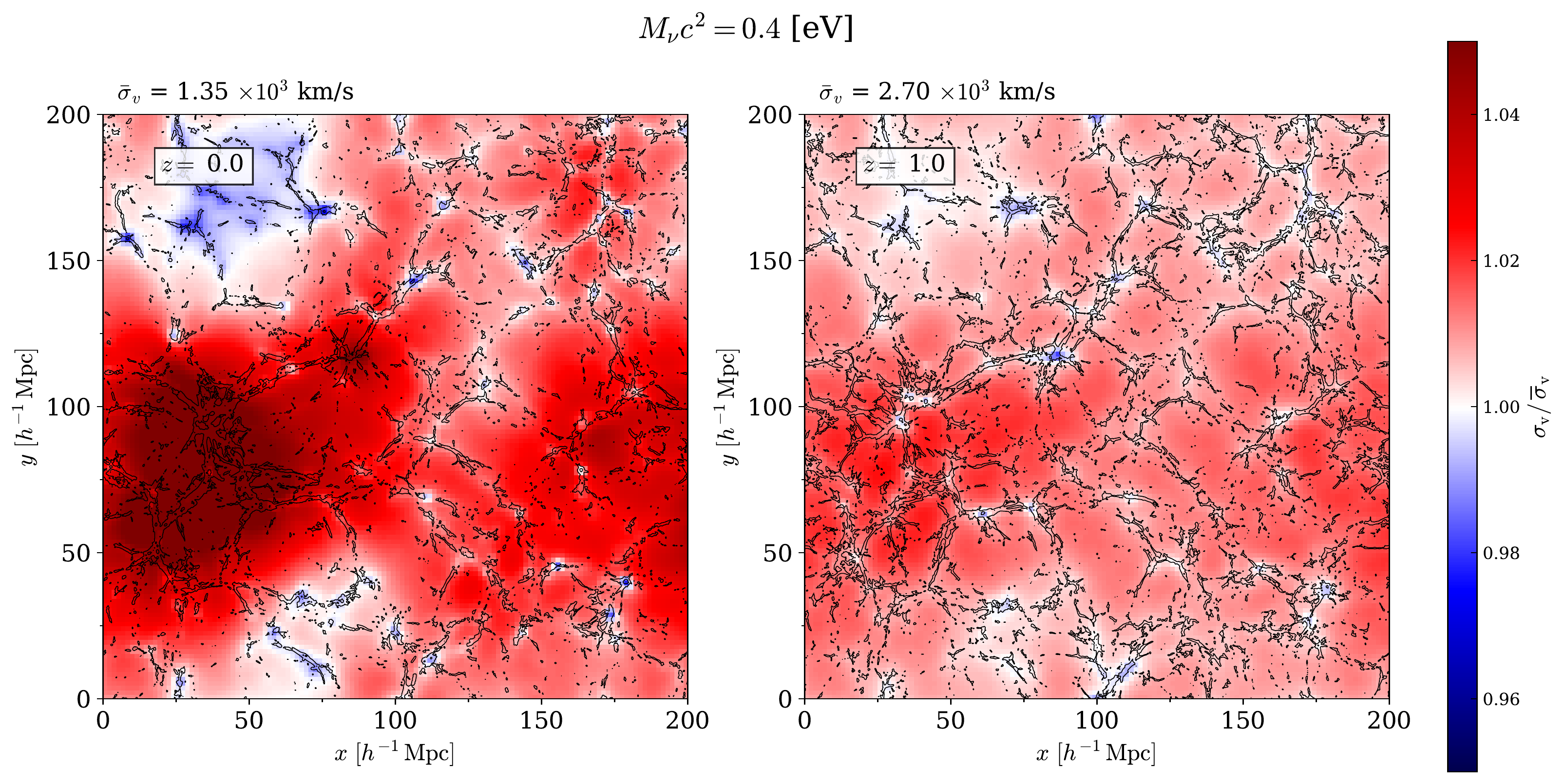}
 \includegraphics[width=17cm]{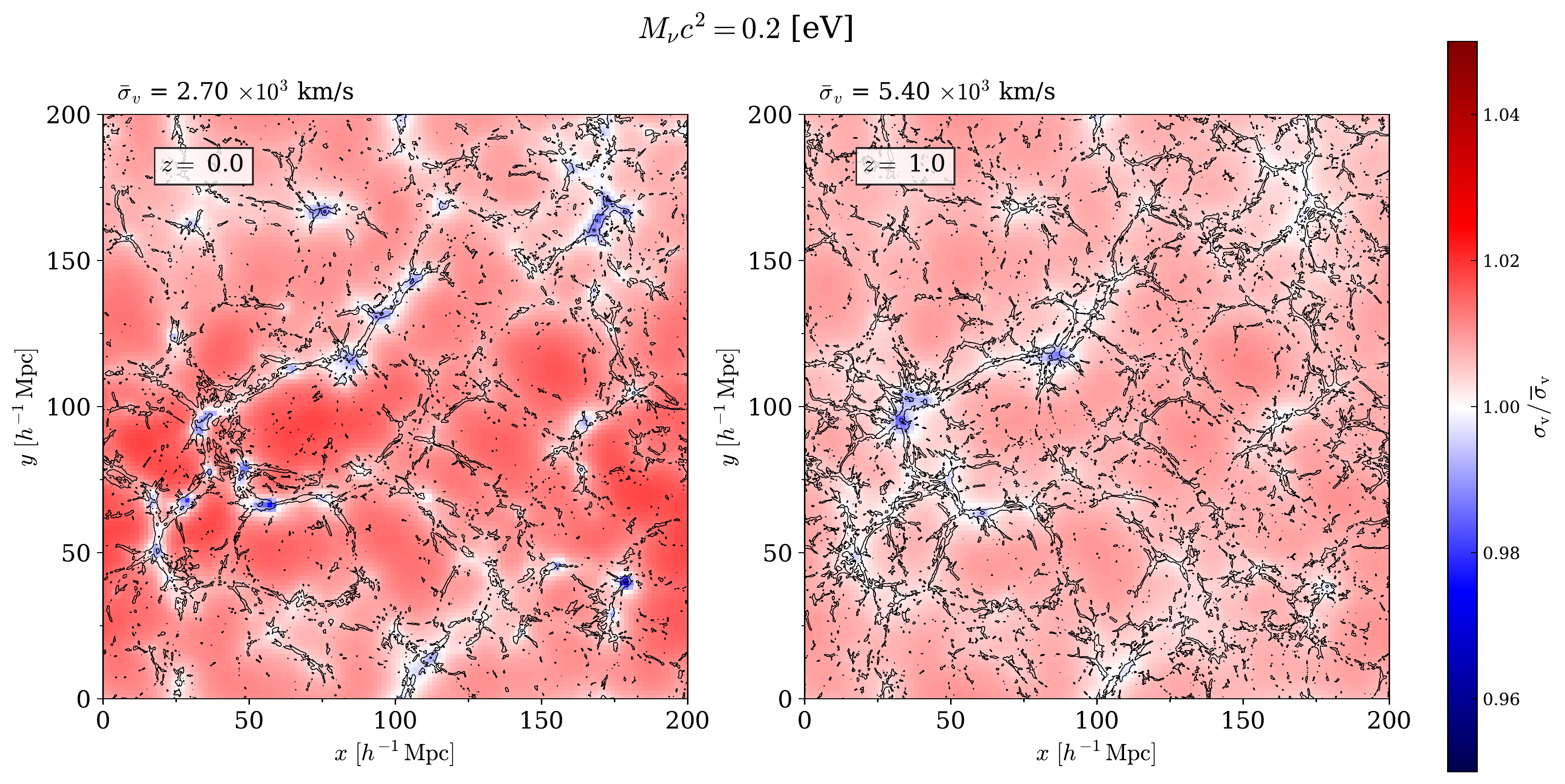}

 \figcaption{The neutrino velocity dispersion map in our runs with
 $M_\nu c^2=0.4$ eV (upper panels) and 0.2 eV (lower panels) at $z=0$
 and $z=1.0$.  Colors indicate the ratio of velocity dispersion with
 respect to the global mean indicated at the left-top of each
 panel. Contours show the mass distribution of CDM.
 \label{fig:snapshot_velcdisp}}
\end{figure*}

Figure~\ref{fig:snapshot_velcdisp} shows the velocity dispersion
$\sigma_{\nu}$ of neutrinos (neutrino "temperature") in the same slice
of the simulation volume as Figure~\ref{fig:snapshot_dens}. Colors
depict the velocity dispersion normalized by the cosmic mean
(Equation~[\ref{eq:velc_disp2}]), the value of which is given in the
top-left to each panel.  In the run with light neutrinos ($M_\nu c^2 =
0.2$ eV), $\sigma_{\nu}$ is lower in and around galaxy clusters, while
we find higher $\sigma_{\nu}$ in void regions. The trend is opposite in
the run with "heavy" neutrinos ($M_\nu c^2 = 0.4$ eV); we find higher
$\sigma_{\nu}$ in higher density regions.

These features can be explained by the fact that the mean velocity
dispersion $\bar{\sigma}_{\rm v}$ ($5.4\times 10^3$ km/s at $z=1$ and
$2.7\times 10^3$ km/s at $z=0$) is significantly larger than typical
virial velocities of DM halos. Neutrinos in the high velocity tail of
the distribution function streams out of DM halos, and then the velocity
dispersion of neutrinos that are trapped within the halos effectively
decreases.  In the run with $M_\nu c^2 = 0.4$ eV, the spatial variation
of velocity dispersion at redshift $z=1$ is similar to that of $M_\nu
c^2 = 0.2$ eV runs at redshift $z=0$. This is because the mean velocity
dispersion is the same between the two cases at the respective epoch.

At $z=0$, the run with $M_\nu c^2 = 0.4$ eV looks significantly
different from the results with $M_\nu c^2 = 0.2$ eV at the same
redshift. This is because the mean velocity dispersion dropped already
to $\sigma_{\rm v} = 1.35 \times 10^3$ km/s, which is comparable to or
smaller than the virial velocities of massive halos (rich galaxy
clusters).  The neutrinos can gravitationally contract into the
potential of the halos and of large-scale filamentary structures. This
gravitational "heating" occurs in the high density regions (Figure
\ref{fig:snapshot_velcdisp}).

\begin{figure*}[htbp]
 \centering
 \includegraphics[width=17cm]{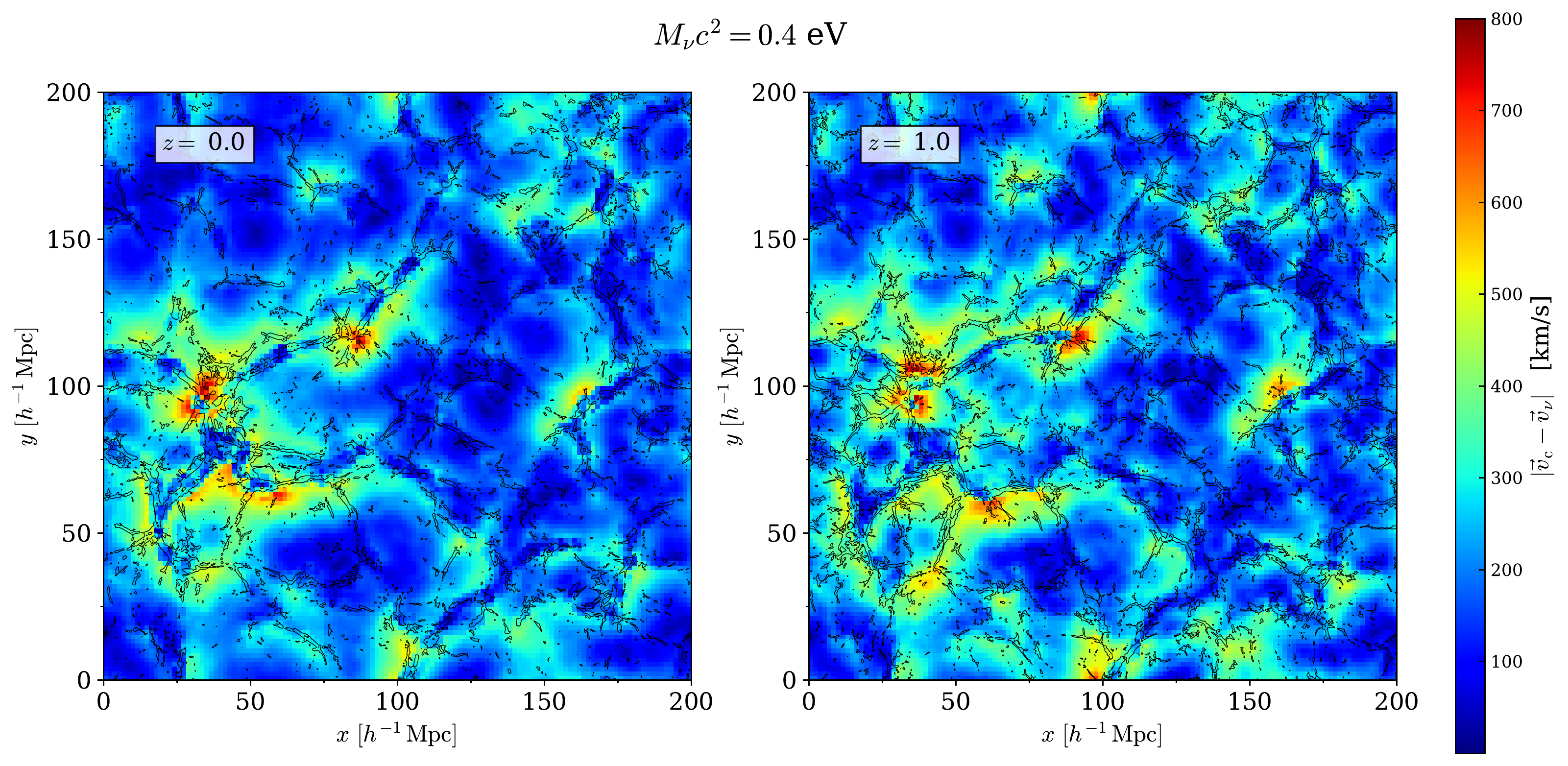}
 \includegraphics[width=17cm]{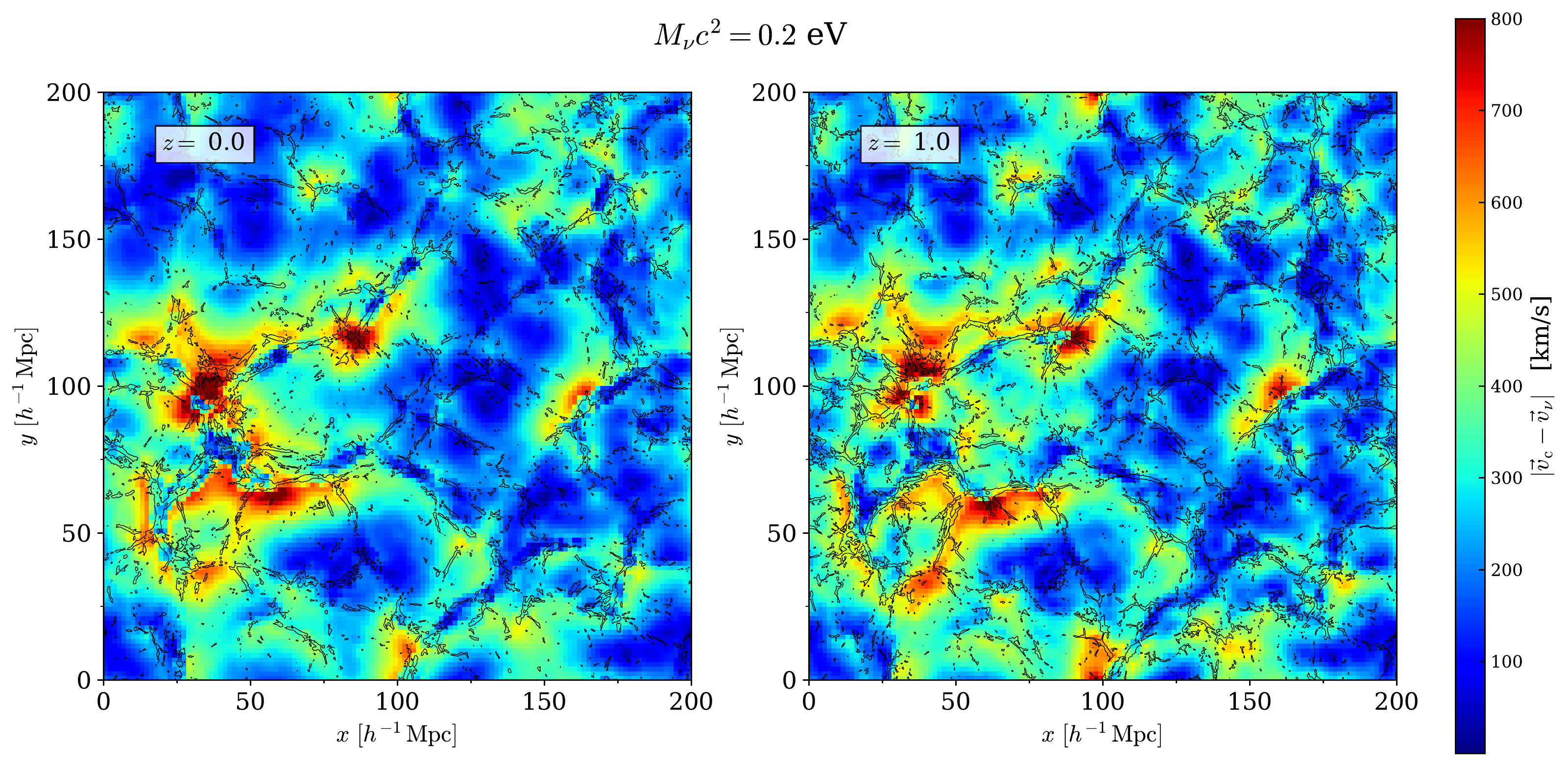}

 \figcaption{Relative velocities between CDM and neutrinos in the same
 regions presented in Figures~\ref{fig:snapshot_dens} and
 \ref{fig:snapshot_velcdisp}. \label{fig:snapshot_velc}}
\end{figure*}

\begin{figure*}[htbp]
 \centering
 \includegraphics[width=18cm]{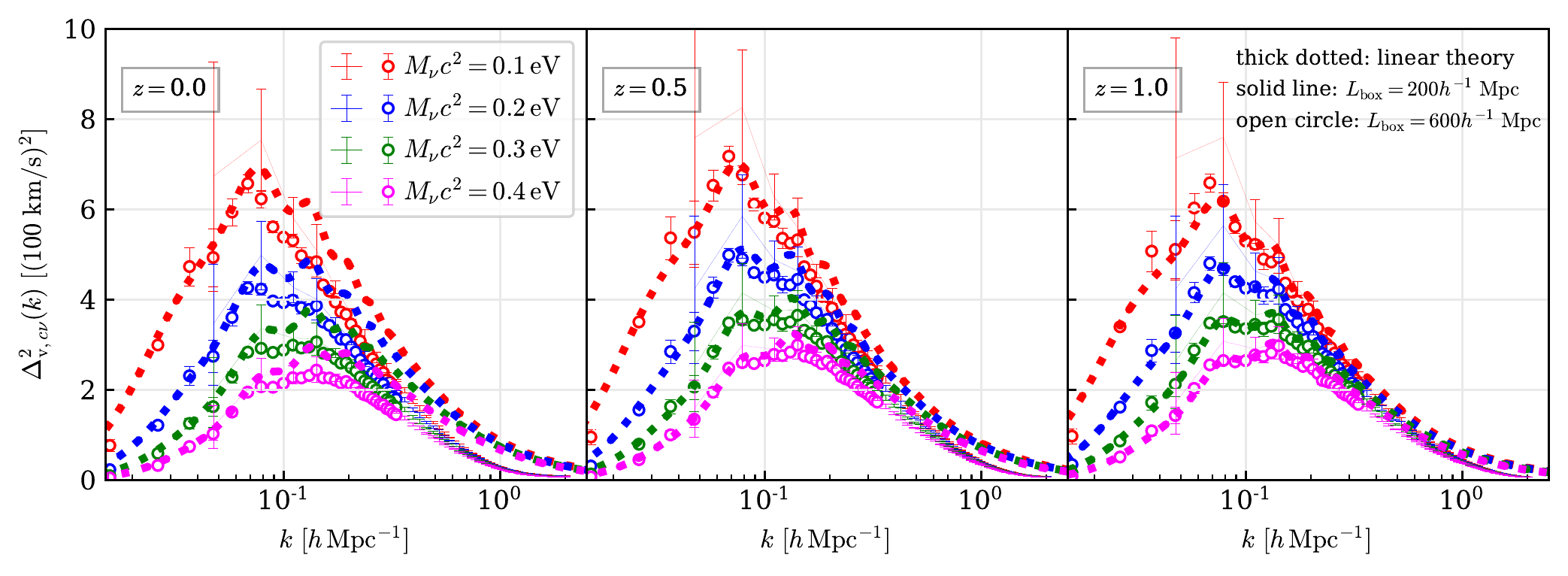}

 \figcaption{Power spectra of relative velocity between CDM and
 neutrinos obtained in the runs with $L_{\rm box}=200 h^{-1}\,{\rm Mpc}$
 (solid lines) and $600h^{-1}\,{\rm Mpc}$ (open circles). Error bars indicate
 the variance among the realizations. Thick dotted lines shows the
 predictions by the linear perturbation theory. \label{fig:vpk}}
\end{figure*}

Figure~\ref{fig:snapshot_velc} shows the relative velocity field between
CDM and neutrinos, $|\itbold{v}_{\rm c}-\itbold{v}_\nu|$, where
$\itbold{v}_{\rm c}$ and $\itbold{v}_\nu$ is the bulk peculiar velocity
of CDM and neutrinos, respectively. We calculate $\itbold{v}_{\rm c}$ by
averaging the peculiar velocities of $N$-body particles residing in the
physical Vlasov mesh grids using the TSC assignment scheme. The neutrino
streaming velocity $\itbold{v}_{\nu}$ is the mean averaged over entire
velocity space.  The relative streaming motions between CDM and
neutrinos induce neutrino wakes in the downstream side of DM halos
\citep{Zhu2014}. The relative motions are primarily driven by the
gravitational clustering of CDM, and the induced "infall" motions of CDM
determine the magnitude. This is manifested by the large relative
velocities in the vicinity of filamentary structures and massive DM
halos as can be seen in Figure~\ref{fig:snapshot_velc}.

On average, the relative velocity magnitude is larger in runs with
smaller neutrino mass, owing to the correspondingly larger neutrino
velocity dispersion.  Less massive neutrinos with large velocities are
not trapped by the local gravitational potential generated by CDM.

We compute the power spectra of the relative velocity given by
\begin{equation}
  \Delta^2_{v,{\rm c\nu}}(k) = \frac{k^3}{2\pi^2}|\itbold{v}_{\rm c}(k)-\itbold{v}_\nu(k)|^2,
\end{equation}
where $\itbold{v}_{\rm c}(k)$ and $\itbold{v}_{\rm \nu}(k)$ are Fourier
transformed peculiar velocities of CDM and neutrinos, respectively. The
computed power spectra are shown in Figure~\ref{fig:vpk}.  The small
amplitudes at $k>1 h {\rm Mpc}^{-1}$ suggest that the relative velocity
is coherent at 10--100 Mpc scales irrespective of neutrino
masses. According to the linear perturbation theory, Fourier transformed
peculiar velocity of a component 's', $\itbold{v}_s(k)$, is given by
\begin{equation}
 \label{eq:linear_continuity}
 \itbold{v}_{\rm s}(k) = -\frac{i\itbold{k}a}{k^2}\dot{\delta}_{\rm s}(k),
\end{equation}
and thus the power spectrum of relative velocity is expressed as
\begin{equation}
 \label{eq:linear_vpk}
 \Delta^2_{v,{\rm c\nu}}(k) = \frac{{\cal P}(k)k^3}{2\pi^2}\left[\frac{\dot{T}_c(k)-\dot{T}_\nu(k)}{k}\right]^2,
\end{equation}
where $T_{\rm c}(k)$ and $T_{\nu}(k)$ are the transfer functions of CDM
and neutrinos, and ${\cal P}(k)$ is the primordial power spectrum of
density perturbation. The linear theoretical prediction for power
spectrum of relative velocity (Equation~(\ref{eq:linear_vpk})) is also
presented in Figure~\ref{fig:vpk} in thick dotted lines, and is roughly
consistent with our numerical results, although the linear theory
predicts slightly higher power spectra than the numerical ones
especially at smaller scales.  \citet{Inman2015, Inman2017} also find
that the simulated relative-velocity power spectra are in good agreement
with linear perturbation theory at large scales with $k\lesssim
0.1\,h/{\rm Mpc}$, but the power amplitude is systematically smaller
than the perturbation theory predictions at nonlinear scales.  Nonlinear
gravitational coupling between CDM and neutrinos may effectively
suppress the velocity offset between the two components.  We note that
the power spectrum measurement is subject to the way of computing the
velocity field of neutrinos from discrete particles \citep{Zhang2015}.

\subsection{Power Spectra of Density Fluctuations}

\begin{figure}[htbp]
 \centering
 \includegraphics[width=8cm]{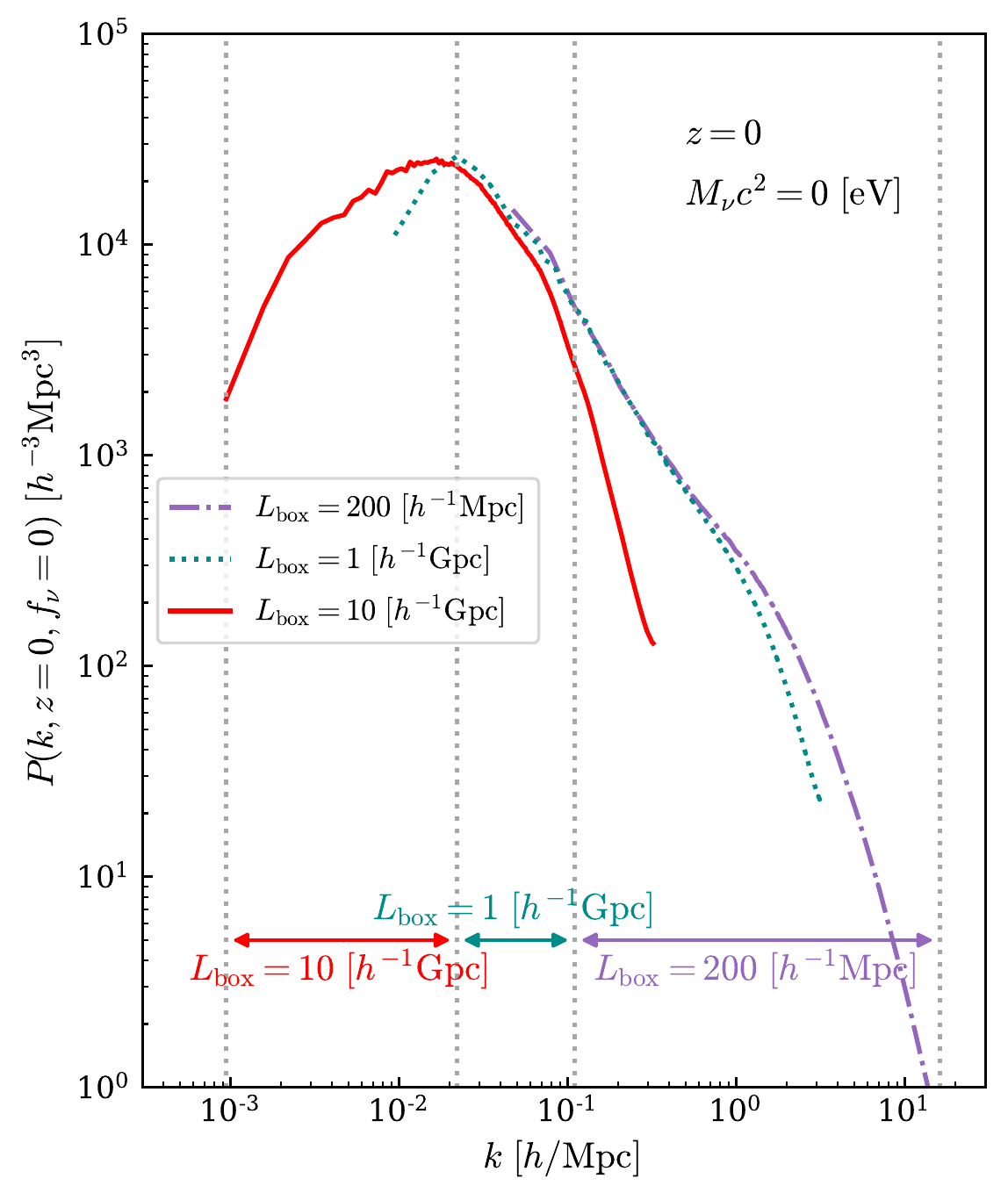}

 \figcaption{Power spectra of CDM density fluctuations obtained from
 runs with $L_{\rm box}=10h^{-1}\,{\rm Gpc}$ (solid), $1h^{-1}\,{\rm
 Gpc}$ (dotted) and $200h^{-1}\,{\rm Mpc}$ (dash-dotted) for the
 massless neutrino cases. At the bottom, we indicate the ranges of wave
 numbers adopted to generate the power spectrum from the simulations
 with different box sizes.  \label{fig:pk} }
\end{figure}

Free-streaming of massive neutrinos causes characteristic suppression of
the amplitude of matter density fluctuations.  We compute the power
spectra of the total matter (CDM + neutrinos) density at a wide range of
wave numbers using our simulation outputs. Figure~\ref{fig:pk} shows the
power spectra of CDM density fluctuation at $z=0$ in the massless case
obtained from the runs with $L_{\rm box} = 10 h^{-1}\,{\rm Gpc}$,
$1h^{-1}\,{\rm Gpc}$ and $200h^{-1}\,{\rm Mpc}$.  For boxsizes of
$L_{\rm box}=200h^{-1}\,{\rm Mpc}$ and $600h^{-1}\,{\rm Mpc}$, we run
four and two realizations and average the power spectra, respectively.
For a given box size of $L_{\rm box}$, the power spectrum is calculated
in the wave number range of $2\pi/L_{\rm box} < k \le k_{\rm ny}$, where
$k_{\rm ny}\equiv\pi N_{\rm PM}^{1/3}/L_{\rm box}$ is the Nyquist
wavelength.  Note that the long wavelength limit ($k\simeq 2\pi/L_{\rm
box}$) of the computed power spectrum is affected by the sample variance
due to the small number of modes, whereas the short wave limit ($k\simeq
k_{\rm ny}$) is influenced by the finite spatial resolution of the
simulations as well as force resolution of the PM $N$-body simulations.
A direct way to check these numerical effects is to compare the power
spectra with different box sizes. For example, the power spectra
obtained from $L_{\rm box}=10h^{-1}{\rm Gpc}$ and $L_{\rm
box}=1h^{-1}{\rm Gpc}$ runs agree with each other at $2\times 10^{-2}
h/{\rm Mpc} < k < 6\times10^{-2}h/{\rm Mpc}$, but we find that the
amplitude at short wavelengths ($k\gtrsim 10^{-1}h/{\rm Mpc}$) is
significantly lower than the ones obtained in the other high resolution
runs.  Considering these features of the power spectra of our
simulations, we decided to construct a synthesized power spectrum by
collecting the power spectra at the respective ``reliable'' wave number
ranges from the results with $L_{\rm box}=10h^{-1}{\rm Gpc}$, $L_{\rm
box}=1h^{-1}{\rm Gpc}$ and $L_{\rm box}=200h^{-1}{\rm Mpc}$ runs.  The
"reliable" ranges are indicated at the bottom of Figure~\ref{fig:pk}.

In order to compute the total matter power spectra contributed both by
CDM and neutrinos, we first compute the CDM mass density by assigning
the particle masses particles onto the regular $N_{\rm PM}=1024^3$ mesh
grids using the TSC mass assignment scheme. We then add the neutrino
density to the mesh grids by up-sampling the neutrino density field with
$N_{\rm x}$ mesh grids.  The obtained total matter density field is
Fourier-transformed to calculate the power spectrum, for which we
correct the aliasing effect due to the finite number of mesh grids
\citep{Jing2005}.

\begin{figure*}[htbp]
\centering
\includegraphics[width=16cm]{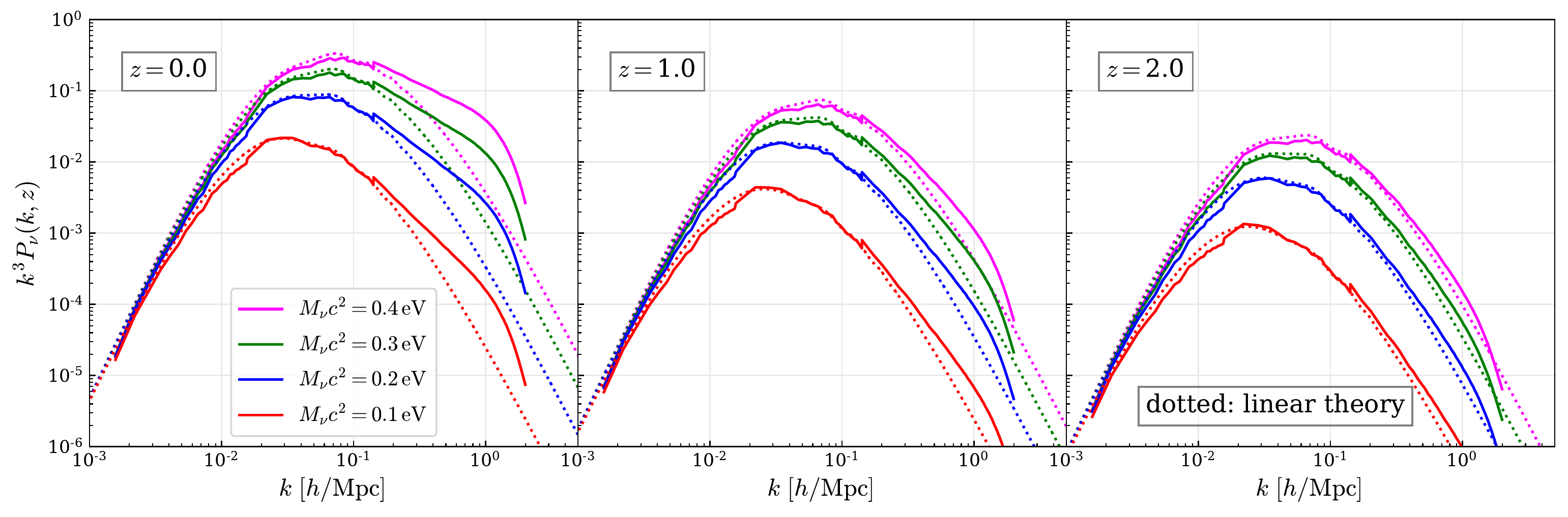}
\figcaption{{\rm Synthesized power spectra of massive neutrinos obtained in our 
simulations at $z=0$, 1 and 2 (solid lines). The dotted lines represent the linear perturbation theory predictions.}\label{fig:pk_nu}}
\end{figure*}

Figure~\ref{fig:pk_nu} shows the dimensionless power spectra of the
neutrino density fluctuations $k^3P_\nu(k)$ calculated from the results
of our $N$-body/Vlasov hybrid simulations, where
$P_\nu(k)=\langle|\delta_\nu(\itbold{k})|^2\rangle$.  Here, we
synthesize the power spectra of the massive neutrinos by combining the
ones obtained by the simulations with $L_{\rm box}=10h^{-1}\,{\rm Gpc}$,
$1h^{-1}\, {\rm Gpc}$ and $200h^{-1}\,{\rm Mpc}$ in the same manner
described above.  Notice the small discontinuous "jumps" at $k\simeq
1.5\times 10^{-1}\,h{\rm Mpc}^{-1}$ and $2\times 10^{-2}\,h{\rm
Mpc}^{-1}$ that are caused by the synthesizing procedure.  Since the
density fluctuations of the neutrino component are intrinsically much
smaller than that of the CDM component, $P_\nu(k)$ can be significantly
contaminated by shot noise, and can even be overwhelmed at small length
scales in particle simulations \citep{Viel2010, Yu2017,
Banerjee2018}. The neutrino density field in our $N$-body/Vlasov hybrid
simulations is reproduced without the shot-noise
contamination. Therefore, we can obtain "clean" power spectra in a
straightforward manner without any additional procedures such as
subtraction of the shot-noise contribution \citep{Yu2017}. We find that
our simulation results are consistent with linear perturbation theory at
$z>2$, but also find nonlinear clustering features at smaller scales of
$k\gtrsim 0.1 h\,{\rm Mpc}$ at later epochs.

\begin{figure*}[htbp]
 \centering
 \includegraphics[width=16cm]{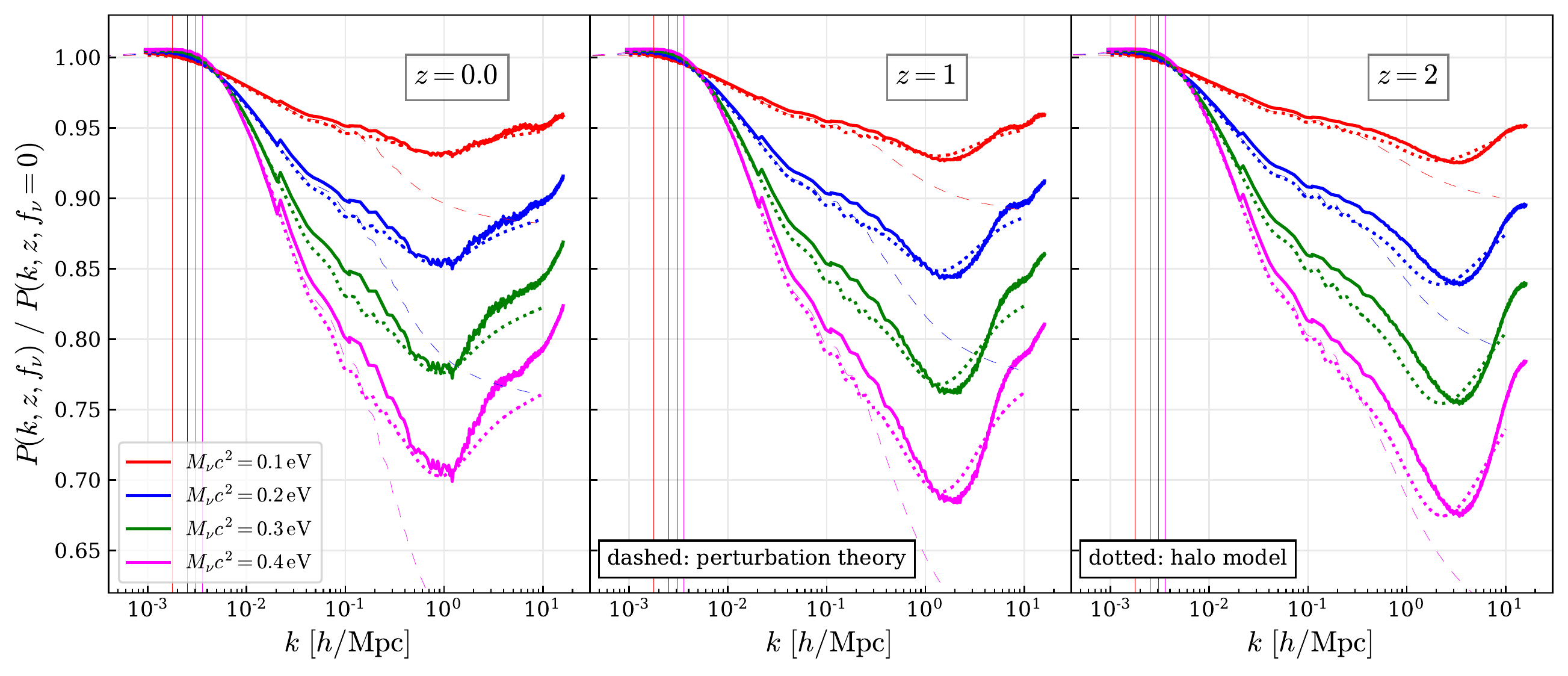}

 \figcaption{Ratios of total matter power spectrum with respect to the
 massless neutrino models at $z= 0, 1$ and $2$. The dashed lines
 represent the results from the 1-loop perturbation theory of
 \citet{Saito2008}, whereas the dashed lines are the halo model
 predictions of \citet{Mead2016}.  The vertical lines at the left of
 each panel indicate the wave numbers corresponding to the maximum free
 streaming scales, i.e., the distance travelled to the epoch of
 non-relativistic transition. \label{fig:pkratio}}
\end{figure*}

Figure~\ref{fig:pkratio} shows the ratios of the total matter power
spectra with respect to those with massless neutrinos at redshifts of
$z=0$, $1$ and $2$ from left to right. We compare the power spectra with
the results of the 1-loop perturbation theory \citep{Saito2008} for
$M_\nu c^2 = 0.1$, 0.2 and 0.4 eV.  The power spectra are consistent at
large length scales of $k < 0.1\,h\,{\rm Mpc}^{-1}$, but the
perturbation theory predicts larger suppression at small scales.  Our
simulations show up-turn features at $k \gtrsim 1\,h\,{\rm Mpc}^{-1}$,
as have been found also in $N$-body simulations
\citep[e.g.][]{Brandbyge2008,Viel2010}.  As depicted in Figure
\ref{fig:pkratio} by dotted lines, the upturn features are in good
agreement with the analytic model of \citet{Mead2016}, which is based on
the standard halo model \citep{Peacock2000, Seljak2000, Cooray2002} with
incorporating the effect of massive neutrinos and calibrated to match
the power spectra obtained in \textsc{cosmic emu} simulations
\citep{Lawrence2010, Heitmann2014} and to simulations by
\citet{Massara2014}.

\begin{figure}[htbp]
 \centering
 \includegraphics[width=8cm]{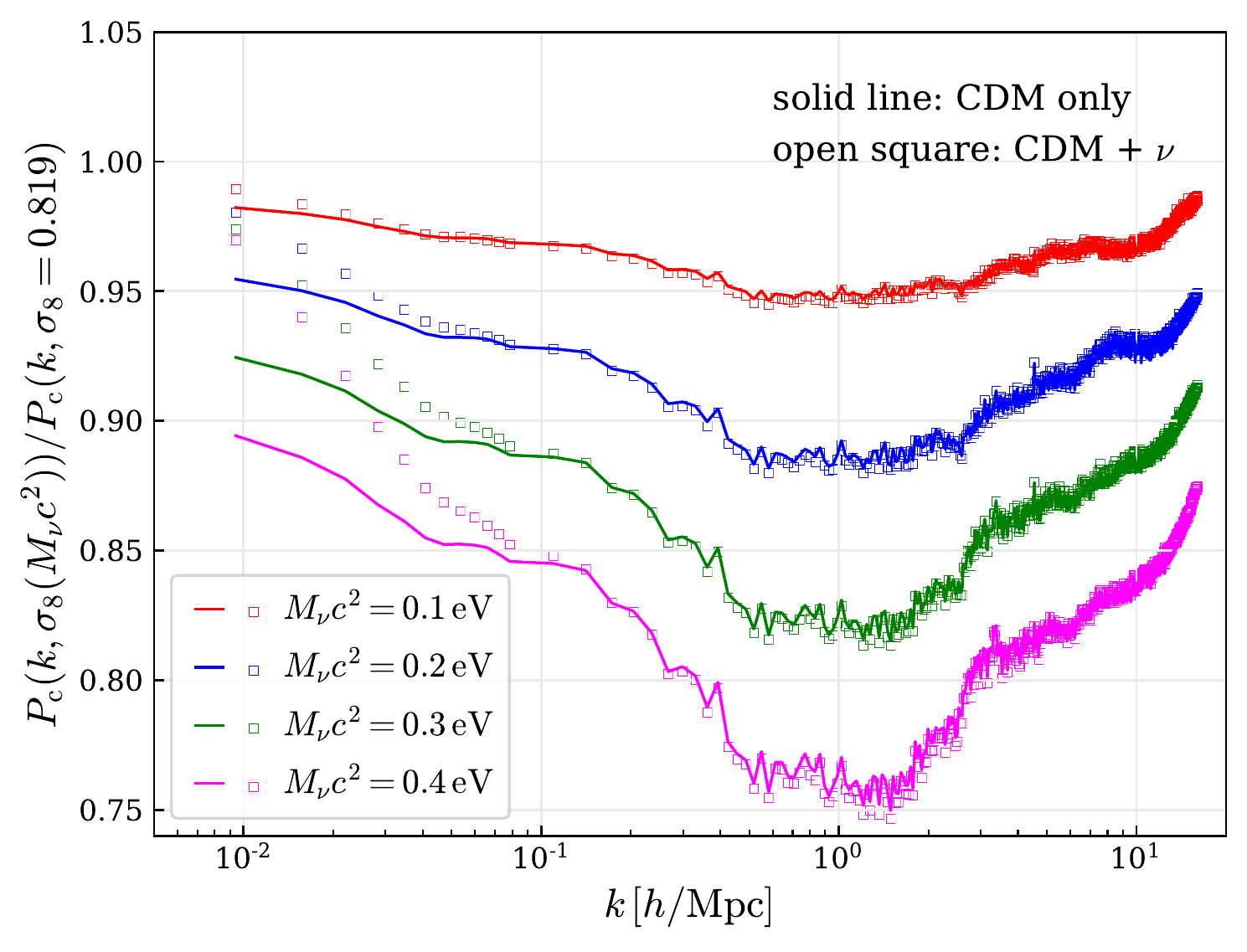}

 \figcaption{Ratios of power spectra of CDM density fluctuation in the
 CDM-only $N$-body simulations with $\Omega_{\rm c}$ and $\sigma_8$
 corresponding to the massive neutrino cases ($M_\nu c^2\neq 0$) with
 respect to the one with $\sigma_8=0.819$ (the massless neutrino case)
 at $z=0$ (solid lines). The ratios of power spectra of CDM density
 fluctuation in the $N$-body/Vlasov hybrid simulation at $z=0$ with
 respect to the one with the massless neutrino case are also presented
 for comparison (open squares).\label{fig:pkratio_wo_nu}}
\end{figure}

In order to examine quantitatively the effect of non-linear CDM
clustering, we perform additional $N$-body simulations with only the CDM
component but with varying cosmological parameters $(\sigma_8,
\Omega_{\rm c})=(0.804, 0.306)$, $(0.785, 0.303)$, $(0.765, 0.301)$ and
$(0.745, 0.299)$ where we also adjust $\Delta_{\cal R}$ at the pivot
scale for a targeted $\sigma_{8}$.  These are the same combinations of
$\sigma_8$ and $\Omega_{\rm c}$ as adopted in the runs with massive
neutrinos listed in Table~\ref{tbl:models}. We compute the ratios of the
CDM power spectra with respect to that with $(\sigma_8,\Omega_{\rm
c})=(0.819, 0.308)$ (the massless neutrino case) and show the results in
Figure~\ref{fig:pkratio_wo_nu}. It is remarkable that, at $k\gtrsim
0.1\,h{\rm Mpc}^{-1}$, the power spectrum ratios of the CDM-only
simulations closely resembles those of the CDM power spectra in
$N$-body/Vlasov hybrid simulations.  We conclude that the up-turn
feature observed in Figure~\ref{fig:pkratio} can be explained by the
difference of nonlinear CDM clustering for different density parameter
$\Omega_{\rm c}$ and the normalization of the density fluctuation
$\sigma_8$. We note that our finding is consistent with the conclusion
of \citet{Massara2014} based on their halo model.

\subsection{Mass Function of DM Halos}
\label{ss:mass_function}

The DM halo mass function is a sensitive statistics to measure/constrain
the neutrino mass.  We identify DM halos in the runs with $L_{\rm
box}=200h^{-1}{\rm Mpc}$ and $600h^{-1}{\rm Mpc}$ using the ROCKSTAR
software package \citep{Behroozi2013}.  We consider DM halos identified
with at least 100 particles. The corresponding halo mass is $9.39\times
10^{10} f_{\rm CDM} M_\odot$ and $2.53\times10^{12} f_{\rm CDM}
M_{\odot}$ for the runs with $L_{\rm box}=200h^{-1}\,{\rm Mpc}$ and
$600h^{-1}\,{\rm Mpc}$, respectively.

\begin{figure*}
 \centering \includegraphics[width=15cm]{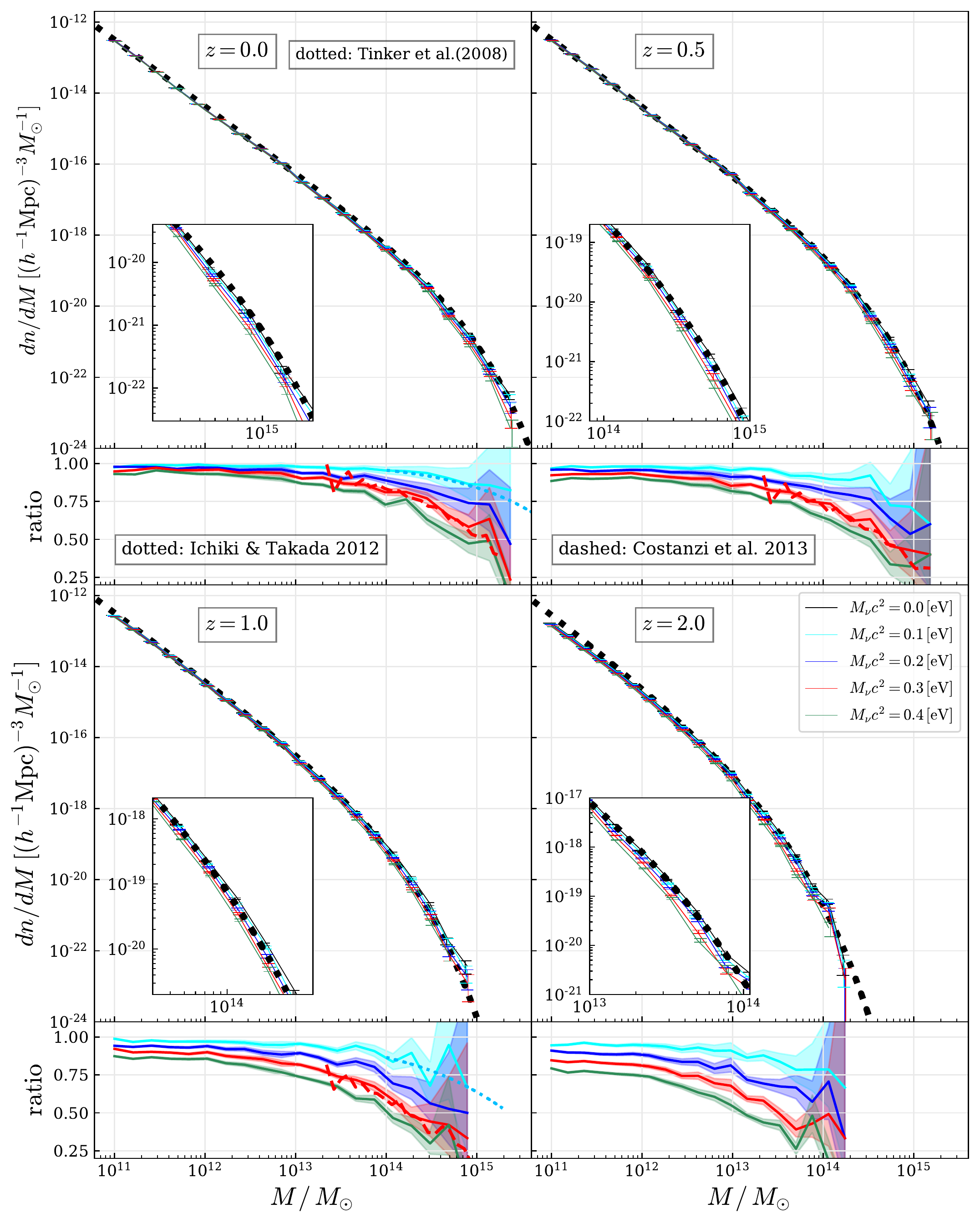}

 \figcaption{Mass functions of DM halos at redshifts of $z=0.0$, 0.5,
 1.0 and 2.0 (from left to right) are shown in upper panels, where
 dotted lines indicate mass functions modeled by
 \citet{Tinker2008}. Error bars indicate the 1-$\sigma$ statistical
 uncertainty.  Lower panels depict ratios of mass functions relative to
 massless neutrino cases, where the shaded regions represent the
 1-$\sigma$ uncertainty around the mean values shown in the solid
 lines. In lower panel, we also plot theoretical predictions given by
 \citet{Ichiki2012} with $M_\nu c^2 = 0.1\,{\rm eV}$ at redshift $z=0$
 and 1 (dotted lines), and numerical results obtained by conventional
 $N$-body simulations \citep{Costanzi2013} with $M_\nu c^2=0.3\,{\rm
 eV}$ at $z=0.0$, 0.5 and 1.0 (dashed lines) for
 comparison. \label{fig:mass_function}}
\end{figure*}

Figure~\ref{fig:mass_function} shows the mass function at $z=0$, 0.5,
1.0 and 2.0.  We also present the ratio of the mass functions relative
to the massless neutrino case in the lower panels.  The mass functions
of the massless cases are in good agreement with the analytic fit of
\citet{Tinker2008}.

There are less massive DM halos in models with massive neutrinos.  We
compare our simulation result for $M_{\nu}c^2 = 0.1$ eV with the model of
\citet{Ichiki2012}.  Their model assumes spherical top-hat collapse in
the presence of massive neutrinos\footnote{ \citet{Ichiki2012} assume
the inverted mass hierarchy in which only two neutrino species have mass
and the other one is massless. This is different from our setting in
which three species have equal masses.}.  The overall good agreement at
$z=0$ suggests that the suppression at large mass scales is caused by
the effective "drag" by free streaming neutrinos.  The mass function for
the $M_\nu c^2=0.3\,{\rm eV}$ model is also consistent with the $N$-body
simulation results of \citet{Costanzi2013}, although the comparison can
be made only at $M > 2\times 10^{13}\,M_{\odot}$. It should be noted
that the mass functions around $M\simeq 10^{11}\,M_{\odot}$ might be
affected by the mild force resolution of the PM scheme adopted in our
$N$-body simulations.

\begin{figure}
 \centering 
 \includegraphics[width=8cm]{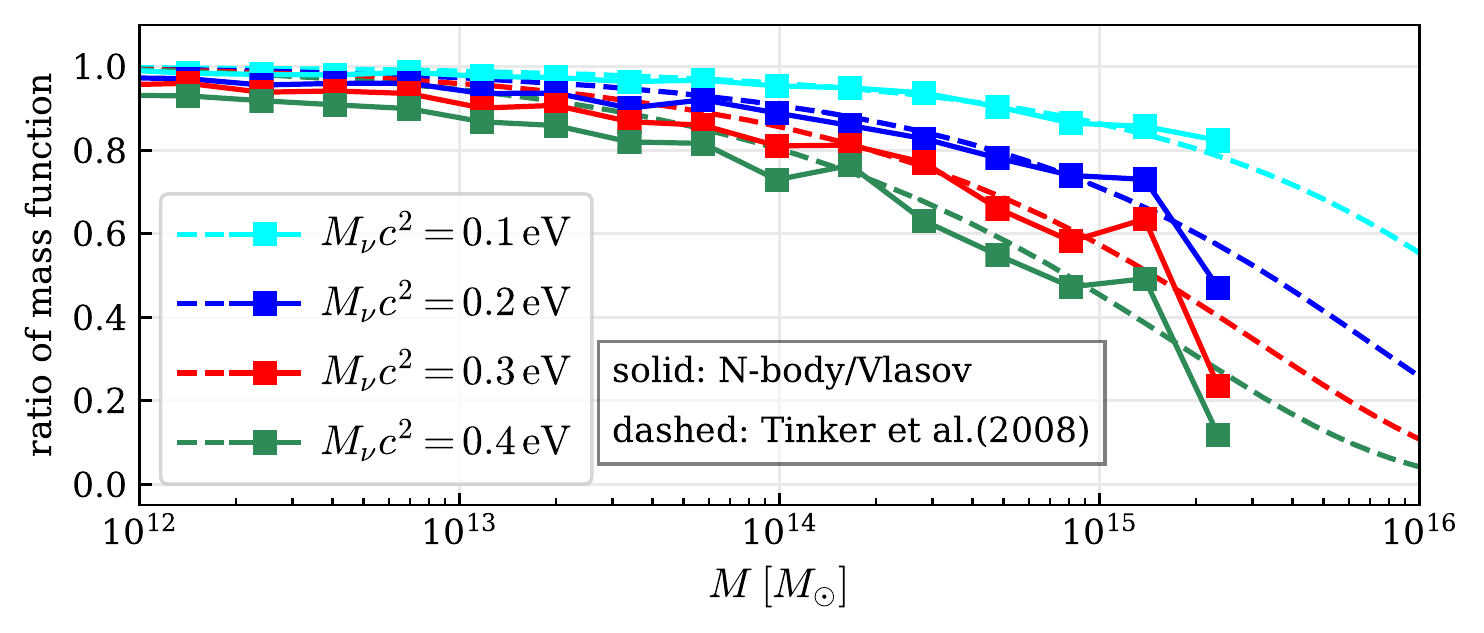}
 
 \figcaption{Ratios of DM halo mass functions obtained in our
 simulations relative to that in the massless neutrino case with
 $(\Omega_{\rm c}, \sigma_8)=(0.308, 0.819)$ (filled squares) are
 compared with those for DM mass functions with $\Omega_{\rm c}$ and
 $\sigma_8$ corresponding to the massive neutrino cases calculated using
 the analytical fit by \citet{Tinker2008} (dashed lines).}
 \label{fig:mf_ratio}
\end{figure}

The fact that the ratios of power spectra of density fluctuation with
respect to the massless neutrino case can be explained by the nonlinear
clustering of the CDM component as shown in
Figure~\ref{fig:pkratio_wo_nu} motivates us to verify whether the same
argument can be also applied to the ratios of DM halo mass functions
presented in Figure~\ref{fig:mass_function}.  Figure~\ref{fig:mf_ratio}
shows ratios of DM halo mass functions with sets of $\sigma_8$ and
$\Omega_{\rm c}$ for massive neutrino cases in Table~\ref{tbl:models}
with respect to the massless neutrino case with $(\sigma_8, \Omega_{\rm
c})=(0.819, 0.308)$ in dashed lines, where the DM halo mass functions
are computed using the analytical fit by \citet{Tinker2008} at redshift
of $z=0$. The ratios are consistent with the ones obtained with our
$N$-body/Vlasov hybrid simulations depicted by the filled squares,
implying that the decrements of massive DM halos in massive neutrino
cases relative to the massless neutrino case also can be explained by
the difference in adopted $\Omega_{\rm c}$ and $\sigma_8$. This result
is consistent with \citet{Castoria2014} in a sense that the universality
of DM halo mass function holds even in the massive neutrino cases, as
long as the rms density fluctuation of the CDM component, rather than
that of the total one, is adopted to compute the DM halo mass function.

\section{Computational Cost\label{sec:cost}}

In this section, we discuss advantages and disadvantages of our
$N$-body/Vlasov hybrid simulations in comparison with conventional
$N$-body simulations in terms of computational cost.  The computational
cost of these numerical simulations includes a variety of quantities
such as CPU/wallclock time, required amount of memory, spatial
resolution, parallelization efficiency and so on.  Since some of these
depend on each other and also on technical properties of the hardwares
on which simulations are performed, it is not very simple to compare
different numerical schemes on different computational
platforms. Therefore, we constrain ourselves to comparison with certain
conditions. Namely, we study (1) the amount of memory needed to achieve
a designated spatial resolution and (2) the computational time compared
with respect to CDM-only simulations with the same number of CDM
particles.

The computational cost of modern cosmological $N$-body simulations
scales roughly as $N\log N$, where $N$ is the number of
particles. $N$-body simulations with massive neutrinos typically set
$N_\nu=N_{\rm c}$ or $N_\nu=8N_{\rm c}$ \citep[e.g.][]{Brandbyge2010b,
Bird2012, Inman2015}, and the computational cost is roughly proportional
to the total numbers of particles because the difference in the
logarithmic factor is small. The memory requirement of $N$-body
simulations is estimated as follows. Sizes for the particle data and the
PM mesh grids are both proportional to the number of particles, while
that for the tree structure is proportional to $\log N$ which has a
small impact to the total memory requirement. Therefore, the overall
memory requirement for a $N$-body simulations is roughly estimated to be
proportional to the total number of particles (CDM+neutrinos).

The computational cost for solving the Vlasov equation with the
directional splitting method
(Equations~[\ref{eq:advection_in_physical_space}] and
[\ref{eq:advection_in_velocity_space}]) is linearly proportional to the
number of mesh grids in six dimension.  Through our numerical
experiments, we find that the CPU time required to integrate the Vlasov
equation is about five times as large as that for the $N$-body
simulation for the CDM component under our settings of the number of
mesh grids ($N_{\rm x}=N_{\rm c}/8^3=128^3$, and $N_{\rm u}=64^3$) and
the number of CDM particles ($N_{\rm c}=1024^3$).  In terms of required
amount of memory, each $N$-body particle has seven words (three for
position, three for velocity, and one for particle index), and thus a
$N$-body simulation needs a total of $7 N_{\rm c}$ words and $N_{\rm c}$
words for the PM mesh grids. The Vlasov part uses $N_{\rm x}N_{\rm u}$
words for the mesh grids.  In our implementation, we adopt single
precision floating-point numbers to store the distribution function in
the Vlasov simulation, and adopt double-precision numbers for particle
data in $N$-body simulations.  Therefore, the total memory size of a
$N$-body/Vlasov hybrid simulation is
\begin{equation}
 8N_{\rm c} + N_{\rm x}N_{\rm u}/2
\end{equation}
in units of a double-precision word.

Table~\ref{tbl:comparison} compares the computational time and required
memory size for our hybrid simulations and conventional $N$-body only
simulations.  We normalize the numbers by the corresponding ones for
CDM-only simulations with the same number of particles.  It should be
noted that the values in the table refer to typical conditions, and they
can vary by a factor of several depending on actual conditions such as
the clustering amplitude of matter and velocity dispersion (or
equivalently, mass) of neutrinos.

The computational time for our hybrid approach is in between those for
$N$-body approach with $N_\nu=N_{\rm c}$ and $N_\nu=8N_{\rm c}$.  The
amount of required memory for the our hybrid approach is even larger
than that of the $N$-body approach with $N_{\nu}=8N_{\rm c}$, but by
only a factor of a few.

Another important issue we have to consider is the spatial resolution of
neutrino distribution and the level of numerical shot-noise. In our
$N$-body/Vlasov hybrid simulations, the spatial resolution is set by the
spatial mesh spacing of Vlasov mesh grids, $L_{\rm box}/N_{\rm
x}^{1/3}$, and the numerical solutions are practically noiseless. On the
other hand, in conventional $N$-body simulations, physical quantities
such as density and velocity fields of neutrinos are obtained by
averaging over a sufficiently large number particles within a certain
averaging length scale.  The number of particles used in the averaging
procedure sets the shot-noise level of the derived physical quantities.
Hence, there is a trade-off between the spatial resolution and the shot
noise in the physical quantities. This is an intrinsic problem with
particle-based simulations.

\floattable
\begin{deluxetable*}{c|c|c|c}
\tabletypesize{\scriptsize}
\tablecaption{Comparison of computational resources and spatial resolution of neutrino distribution with conventional $N$-body simulations.\label{tbl:comparison}}
\tablehead{
\colhead{numerical scheme} &\colhead{computational time} & \colhead{memory size} & \colhead{spatial resolution of neutrino}
}

\startdata
$N$-body w/o neutrinos & 1 & 1 & --\\
\hline
\begin{tabular}{c}
 $N$-body+Vlasov \\
 ($N_{\rm x} = N_{\rm c}/8^3$, $N_{\rm u}=64^3$)
\end{tabular} 
& $\simeq 1+5\displaystyle\left(\frac{8^3 N_{\rm x}}{N_{\rm c}}\right)\left(\frac{N_{\rm u}}{64^3}\right)\left(\frac{\log 1024^3}{\log N_{\rm c}}\right)$ & $1+32\displaystyle\left(\frac{8^3 N_{\rm x}}{N_{\rm c}}\right)\left(\frac{N_{\rm u}}{64^3}\right)$ & $\displaystyle\frac{L_{\rm box}}{N_{\rm x}^{1/3}}=8\left(\frac{N_{\rm c}^{1/3}}{8N_{\rm x}^{1/3}}\right)\frac{L_{\rm box}}{N_{\rm c}^{1/3}}$ \\
\hline
$N$-body    ($N_{\nu} = N_{\rm c}$)  & $\displaystyle \left(1+\frac{N_\nu}{N_{\rm c}}\right)\frac{\log (N_{\rm c}+N_\nu)}{\log N_{\rm c}}\gtrsim 2$ & $\displaystyle\gtrsim 1+\frac{N_\nu}{N_{\rm c}}=2$ &$\displaystyle 4\left(\frac{n_{\rm a}}{64}\right)^{1/3}\left(\frac{N_{\rm c}}{N_\nu}\right)^{1/3}\frac{L_{\rm box}}{N_{\rm c}^{1/3}}$\\
\hline
$N$-body    ($N_{\nu} = 8N_{\rm c}$) & $\displaystyle \left(1+\frac{N_\nu}{N_{\rm c}}\right)\frac{\log (N_{\rm c}+N_\nu)}{\log N_{\rm c}}\gtrsim 9$ & $\displaystyle\gtrsim 1+\frac{N_\nu}{N_{\rm c}}=9$ &$\displaystyle 2\left(\frac{n_{\rm a}}{64}\right)^{1/3}\left(\frac{8N_{\rm c}}{N_\nu}\right)^{1/3}\frac{L_{\rm box}}{N_{\rm c}^{1/3}}$
\enddata 
\end{deluxetable*}

Unlike the strong clustering of CDM, the density fluctuation of
neutrinos remains relatively small from the early universe through to
the present epoch.  When averaging over $n_{\rm a}$ $N$-body particles,
the smoothing scale is approximately estimated to be $n_{\rm a}^{1/3}$
times mean inter-particle separation $L_{\rm box}/N_{\nu}^{1/3}$, and
the fractional uncertainty of the averaged quantities is estimated to be
$n_{\rm a}^{-1/2}$ assuming the Poissonian statistics. Therefore, if we
adopt $n_{\rm a}=64$, the spatial resolution of the neutrino component
is 4 times the mean inter-particle separation. The fractional shot-noise
level is then $\simeq$ 12.5\%. In Table~\ref{tbl:comparison}, we compare
the effective spatial resolution of our $N$-body/Vlasov hybrid
simulations and of conventional $N$-body simulations.  With the current
setting of $N_{\rm c}$, and $N_{\rm x}$, the spatial resolution of our
hybrid simulations is eight times the mean separation of the CDM
particles.  The spatial resolution in conventional $N$-body simulations
is four and two times the mean separation of CDM particles for
$N_\nu=N_{\rm c}$ and $N_\nu=8N_{\rm c}$, respectively, which is better
than that in our hybrid approach, but the level of shot noise leaves
$\simeq 10$\% uncertainty.

\bigskip

\section{Summary and Discussion\label{sec:summary}}

We have performed a set of cosmological simulations of the large-scale
structure formation with massive neutrinos.  We solve the Vlasov-Poisson
equations to follow the time evolution of the distribution function of
neutrinos in the six-dimensional phase space.  Our simulations are able
to reproduce accurately kinematic phenomena that strongly depend on the
velocity distribution, such as free-streaming and collisionless damping
of neutrinos and their wakes produced by the gravitational interaction
with dark matter halos.

We use the simulations to study the effect of cosmological relic
neutrinos on structure formation by varying the neutrino mass. We find
that the neutrinos "condensate" onto the large-scale structure
\citep{Yu2017}.  The neutrino mass density varies significantly across
the cosmological volume, exhibiting nonlinear clustering.  The
neutrino-rich regions are strongly correlated with massive DM halos
(galaxy clusters). We also find that the neutrino velocity dispersion
$\sigma_{\nu}$, or the effective "temperature", differs significantly
depending on the neutrino mass.  Interestingly, $\sigma_{\nu}$ can be
smaller than the cosmic mean in DM halos and filaments.  This manifest
selective escape of neutrinos in the high velocity tails of the
distribution function, which leaves {\it cooler} neutrinos inside
virialized objects.

We find larger relative streaming velocities in the case with lighter
neutrinos because less massive neutrinos have a larger velocity
dispersion whereas their bulk velocities are insensitive to local
gravitational potential field. The resulting power spectra of the
relative velocity field is consistent with that predicted by linear
perturbation theory at large length scales of $k\lesssim 0.1 h/{\rm
Mpc}$. The power amplitudes at smaller scales are significantly smaller
than the linear theory prediction, suggesting nonlinear density-velocity
coupling.  These results are consistent with \citet{Inman2015,
Inman2017}.

The dynamical effect of massive neutrinos on the LSS formation is
investigated in terms of damping of matter power spectra and the
abundance of massive DM halos. The matter power spectra are found to be
in good agreement with the first-order perturbation theory prediction of
\citet{Saito2008} at wave numbers of $k<0.1\, h/{\rm Mpc}$. Our
simulation results exhibit departure from the prediction at smaller
scales.  We also find that such nonlinear features in matter power
spectra are well reproduced by the halo model by \citet{Mead2016} and
can be explained by the non-linear clustering of the CDM component.  Our
$N$-body/Vlasov hybrid simulations enable us to compute the power
spectra of the neutrino component without numerical shot noise. We find
that the power spectra are consistent with the prediction of linear
perturbation theory at high redshift ($z>2$), but show significant
nonlinear evolution at $k\gtrsim 0.1\,h{\rm Mpc}$ at lower redshift.

As for the halo abundance, it is found that the mass function of DM halos
is decreased under the presence of massive neutrinos and that the
decrement is larger for more massive DM halos and for larger neutrino
masses and is found to be consistent with previous numerical simulation
by \citet{Costanzi2013} and also with the analytical prediction based on
the spherical top-hat collapse model in the mixture of CDM and massive
neutrinos \citep{Ichiki2012}.

Overall, the results obtained by our $N$-body/Vlasov hybrid simulations
agree well with those of $N$-body only simulations.  It is encouraging
that the two, completely different approaches to reproduce the dynamics
of massive neutrinos in the LSS formation yield essentially consistent
results. Conventional $N$-body simulations are not severely compromised
by the shot-nose due to coarse particle sampling as long as the
statistical quantities of the dark matter density field and of dark
halos are concerned.  Our $N$-body/Vlasov hybrid approach has an
advantage in reproducing kinematic phenomena in which the phase-space
structure of neutrinos directly affect the dynamics. Generation of
neutrino wakes is one such example.  We continue exploring the nonlinear
dynamics of collisionless systems such as neutrinos and hot/warm dark
matter.

\bigskip\bigskip

We thank the anonymous referee for his/her valuable comments on the
earlier drafts of the manuscript. We also thank Shunsuke Hozumi for
fruitful discussion in the early stage of this project.  This research
is supported by the Grant-in-Aid for Scientific Research by the JSPS
KAKENHI Grant Number JP18H04336 and by MEXT as ``Priority Issue on
Post-K computer'' (Elucidation of the Fundamental Laws and Evolution of
the Universe) and as ``Program for Promoting Researches on the
Supercomputer Fugaku'' (Toward a unified view of the universe: from
large scale structures to planets).  ST acknowledges support from JST
AIP Acceleration Research Grant Number JP 20317829.  Numerical
simulations presented in this paper are performed on computational
resources of the K computer provided by the RIKEN Center for
Computational Sciences through the HPCI System Research project (project
ID:hp170231, hp180049 and hp190161). This research also uses
computational resources of the Oakforest–PACS through the HPCI System
Research Project (project ID:hp170123 and hp190093) and
Multidisciplinary Cooperative Research Program in Center for
Computational Sciences, University of Tsukuba.

\bibliography{ms}

\appendix

\section{Choice of velocity coordinate in cosmological Vlasov equation}
\label{sec:velocitycoordinate}

In the main text, we adopt the cosmological Vlasov equation in the form
of Equation~(\ref{eq:comoving_vlasov}) where the canonical velocity
$\itbold{u}=a^2 \dot{\itbold{x}}$ is used.  One can also choose the
peculiar velocity $\itbold{v}=a\dot{\itbold{x}}$ as a velocity
coordinate of the Vlasov equation, which can then be re-written as
\begin{equation}
 \label{eq:comoving_vlasov_2}
 \pdif{\hat{f}}{t} + \frac{\itbold{v}}{a}\cdot\pdif{\hat{f}}{\itbold{x}} -\left[H\itbold{v}+\frac{\nabla\phi}{a^2}\right]\cdot\pdif{\hat{f}}{\itbold{v}} = 0,
\end{equation}
where $\hat{f}(\itbold{x},\itbold{v},t)$ is the distribution function as
a function of $\itbold{x}$, $\itbold{v}$ and $t$. It is related to
$f(\itbold{x}, \itbold{u}, t)$ in Equation~(\ref{eq:comoving_vlasov}) as
$\hat{f}(\itbold{x}, \itbold{v}, t) = af(\itbold{x}, \itbold{u}, t)$.

\bigskip

Although Equations~(\ref{eq:comoving_vlasov}) and
(\ref{eq:comoving_vlasov_2}) are physically equivalent with each other,
the choice of velocity coordinate is of critical importance in numerical
simulations. Since we adopt a finite volume method to solve the Vlasov
equation, we need to determine the extent of velocity space {\it before}
running a simulation so that the pre-defined velocity space covers the
actual velocity distribution realized in the numerical simulation. In
the light of this, Equation~(\ref{eq:comoving_vlasov}) is suitable for
simulations with massive neutrinos, because their velocity dispersion is
much larger than their typical bulk velocities, and because the
dispersion of canonical velocity $\itbold{u}=a^2\dot{\itbold{x}}$
remains roughly constant during the initial, linear evolution phase.

\bigskip

 CDM or warm dark matter (WDM); their bulk velocity is much larger than
their thermal velocity dispersion contrary to the case with massive
neutrinos, matter distribution in the canonical velocity space quickly
changes due to the quadratic dependence on the scale factor $a(t)$, and
eventually exceeds the pre-determined extent of the numerical
``velocity'' space. We find that Vlasov simulations of CDM and WDM are
successfully performed by numerically integrating
Equation~(\ref{eq:comoving_vlasov_2}), and should be presented in our
future works.

\end{document}